\documentstyle[12pt,a4wide,epsf]{article}

\renewcommand{\theequation}{\arabic{section}.\arabic{equation}}

\begin{document}

\begin{titlepage}
\renewcommand{\thefootnote}{\fnsymbol{footnote}}
\makebox[2cm]{}\\[-1in]
\begin{flushright}
\begin{tabular}{l}
TUM--T31--67/94/R\\
MPI--PhT/94--49/R\\
UAB--FT--347/R\\
hep--ph/9408306
\end{tabular}
\end{flushright}
\vskip0.4cm
\begin{center}
{\Large\bf Charm Quark Mass Dependence of QCD Corrections\\[4pt]
to Nonleptonic Inclusive B Decays}

\vspace{2cm}

E.\ Bagan$^1$\footnote{Permanent address: Grup de F\'\i sica
Te\`orica, Dept.\ de F\'\i sica and Institut de F\'\i sica d'Altes
Energies, IFAE, Universitat Aut\`onoma de Barcelona,
E--08193 Bellaterra (Barcelona), Spain},
Patricia Ball$^2$\footnote{Address after September 1994: CERN, Theory
Division, CH--1211 Gen\`{e}ve 23, Switzerland},
V.M.\ Braun$^3$\footnote{Address after August 1994: DESY, Notkestra\ss\/e
85, D--22603 Hamburg, Germany} and
 P. Gosdzinsky$^4$

\vspace{1.5cm}

$^1${\em Physics Department, Brookhaven National Laboratory, Upton, NY
11973, USA}\\[0.5cm]
$^2${\em Physik--Department/T30, TU M\"{u}nchen, D--85747 Garching,
Germany}\\[0.5cm]
$^3${\em Max--Planck--Institut f\"{u}r Physik, P.O.\ Box 40 12 12,
D--80805 M\"{u}nchen, Germany}\\[0.5cm]
$^4${\em Grup de F\'\i sica Te\`orica, Dept. de F\'\i sica and
Institut de F\'\i sica d'Altes Energies, IFAE,
Universitat Aut\`onoma de Barcelona,
E--08193 Bellaterra (Barcelona), Spain}

\vspace{1cm}

{\em September 6, 1994}

\vspace{1cm}

{\bf Abstract:\\[5pt]}
\parbox[t]{\textwidth}{
We calculate the radiative corrections to the nonleptonic inclusive
B decay mode $b\rightarrow c\bar u d$ taking into account the
charm quark mass. Compared to the massless case, corrections resulting from
a nonvanishing c quark mass increase the nonleptonic rate by (4--8)\%,
depending on the renormalization point. As a by--product of our
calculation, we obtain an analytic expression
 for the radiative correction to the
semileptonic decay
$b\rightarrow u\tau\bar\nu$ taking into account the $\tau$ lepton mass,
and estimate the c quark mass effects on the nonleptonic
decay mode $b\rightarrow c\bar c s$.
}

\vspace{1cm}
{\em To appear in Nucl.\ Phys.\ B}
\end{center}
\end{titlepage}
\setcounter{footnote}{0}
\newpage

\section{Introduction}
The past few years have witnessed a continuous interest in the physics of
heavy quarks, in particular B meson decays. In addition to their
potential r\^{o}le as precision tests of the Standard Model, they also
provide important information on the Cabibbo--Kobayashi--Maskawa (CKM)
matrix elements $V_{cb}$, $V_{bu}$, and others. From a theoretical
point of view, inclusive widths tend generally to be cleaner than
exclusive ones. This is in particular true for the decays of heavy
hadrons, which are essentially short distance processes since the
amount of energy released is large compared with a typical confinement
scale of $\sim 1\,$GeV. Indeed it can be shown, cf.\
\cite{general,nonlep},
that up to corrections of order $1/m_b^2$, $m_b$ being the b quark
mass, inclusive decay rates of beautiful hadrons are essentially the same
as for the underlying quark processes and thus can reliably be
calculated in perturbation theory. Moreover, the corrections suppressed
by powers of the heavy quark mass can be evaluated in a
systematic way, using the operator product expansion (OPE) technique
combined with the heavy quark expansion \cite{general,nonlep}.
At present, it is widely believed that
inclusive decay widths --- semileptonic, rare radiative or
hadronic --- may be reliably predicted to an accuracy limited mainly
by the unavoidable shortcomings of perturbative calculations such as
scale-- and scheme--dependence.
There is little controversy that these calculations rest on a firm
theoretical foundation.

A high accuracy thus requires the calculation of radiative
corrections to inclusive widths, a task pioneered by the authors of
\cite{CM78}, where the radiative correction to the semileptonic decay
width of the D meson was obtained from earlier calculations \cite{old} of
the muon decay in QED. At present, ${\cal O}(\alpha_s)$ corrections
are known for both the semileptonic and the hadronic
decay width \cite{ACMP81,GB93}, in the approximation of massless final state
quarks.

Taking into account the masses of the final state particles has proved
to be a difficult task. A significant effort in this direction has been
made by Hokim and Pham \cite{HP}, who calculated one--gluon corrections
to inclusive weak decays taking full account of all the masses.
Nir \cite{Nir89} has further found an analytical expression for the radiative
correction to the semileptonic $b\to c e\bar\nu $ decay,
 the numerical results being given in
\cite{CM78}. Unfortunately, the results of \cite{HP} are not sufficient
for the calculation of inclusive widths with renormalization group
improvement to two--loop accuracy, since in addition to the
$\alpha_s$ correction calculated by Hokim and Pham one must take into
account terms proportional to $\alpha_s^2\ln M_W^2/m_b^2$ which are
equally important and are also in general mass--dependent.
Thus, in  practice of data analysis quark mass corrections
to nonleptonic decays are generally being omitted. On the other hand,
it is known that the effect of non--zero masses on the radiative
corrections can be very large, especially if they involve
colour--attraction in the final state. An example was provided
recently within the heavy quark effective theory: radiative
corrections to correlation functions, containing one heavy and
one light quark and evaluated near the mass--shell of the heavy quark,
turned out to be by almost an order of magnitude
larger than for light quarks \cite{BG92}. Thus
the actual size of radiative corrections in nonleptonic
decays is essentially not fixed and deserves a detailed
study. We address this problem in the present paper.

Apart from the pursuit of completeness of formulas,
our study has been fuelled by an
intriguing conflict of theory with experiment, concerning the
semileptonic branching ratio of the B meson:
over the last few years the measured
semileptonic branching ratio \cite{ARGUS}
has consistently turned out to be (15--20)\% smaller than
theoretical expectations, and the reason for this discrepancy is
not understood. This problem has attracted already considerable
attention among theorists \cite{AP91,PS93,BBSV941,FWD94}. We
believe that the knowledge of radiative corrections
to nonleptonic decays with full account for the c quark mass
is imperative for a quantitative understanding of the problem.

A principal new result of this paper is the calculation of the
QCD radiative correction to the inclusive decay mode
\begin{itemize}
 \item $ b \to c \bar u d $
\end{itemize}
including full dependence on the c quark mass and the renormalization
group improvement to the two--loop accuracy.
This decay dominates the hadronic B meson width
and is crucial for understanding the semileptonic branching ratio.
We complete the
calculations of \cite{HP} by studying relevant
corrections due to operator--mixing and  obtain analytic expressions
for all radiative corrections with one massive quark in the final state,
which so far have partly been only available in form of
one--dimensional parametric integrals \cite{HP}.
As a by--product of this analysis, we obtain an analytic expression for
radiative corrections to the decay
\begin{itemize}
 \item $ b \to u \tau \bar\nu $
\end{itemize}
taking into account the $\tau$ lepton mass.
Decays involving the $\tau$ leptons are potentially very interesting,
since they may be used to constrain certain extensions of the Standard
Model and attract an ever increasing experimental and theoretical
interest \cite{tau}.

In addition, we estimate QCD corrections to
the decay
\begin{itemize}
 \item $ b \to c \bar c s $ \,,
\end{itemize}
considering the colour--favoured contribution,
and argue that in this case the radiative correction is
 strongly enhanced when the c quark mass is included.
When appropriate, we compare our results with those of Hokim and Pham.

With realistic values of the b and c quark masses, the nonleptonic
rate $b \to c \bar u d $ is increased by approximately
(4--8)\%, depending on the renormalization scale, if a non--zero
c quark mass is taken into account which has to be compared
with a (4--7)\% increase for the semileptonic $b \to c e\bar\nu $
rate. Thus, the ratio $\Gamma(b\to c ud)/\Gamma(b\to c e \bar\nu)$
is only weakly affected by taking into account the c quark mass
in the radiative corrections. This compensation is due to the specific
structure of the effective weak Lagrangian, which supresses the gluon
exchange between the c and the light quarks.
The $\tau$ lepton mass corrections to the decay
$ b \to u \tau \bar\nu $ turn out to be of the same
order of magnitude as c quark mass corrections  to the
semileptonic $b \to c e \bar\nu$ decay.

Although the calculations are straightforward in principle,
in practice they turn out to be
 rather tedious and involve several delicate points which we
discuss in detail below. We use a conservative technique and
calculate total imaginary parts by means of Cutkosky rules, i.e.\
we sum up contributions corresponding to real and virtual gluon
emission, and introduce small masses for the gluon and the
light quark to avoid infra--red (IR) divergences due to the presence
of massless
particles in the final state. The virtual corrections are calculated
using dimensional regularization, a framework that requires special
care with the definition of the $\gamma_5$ matrix.
We use the so--called na\"{\i}ve dimensional regularization (NDR) scheme,
in which $\gamma_5$ anticommutes in $D$ dimensions. This scheme has
been shown to yield correct results for the two--loop
anomalous dimensions in the relevant terms of the
effective weak Lagrangian \cite{BW90}.
In the limit of zero masses of all final state particles
our results coincide with those of Altarelli et al., \cite{ACMP81},
obtained in the dimensional reduction scheme, and with those of
Buchalla, \cite{GB93}, obtained in the 't Hooft--Veltman scheme.

Our paper is organized as follows: in Sec.~2 we recall the general
framework for the calculation of weak processes to next--to--leading
order in QCD, list the necessary contributions to be calculated,
explain the particular calculation procedure which we use throughout
this paper, and discuss various subtle points.
Sec.~3 contains our main results: the analytic expressions for the
radiative corrections for non--zero c quark ($\tau$ lepton) mass,
along with simple parametrizations. In Sec.~4 we estimate the
effect of finite mass corrections on the decay $b\to c\bar c s$.
Sec.~5 contains the numerical analysis and a discussion of various
phenomenological applications. Sec.~6 completes the paper with a
summary and conclusions.
Technical details of the calculation are given in the appendices.

\section{Theoretical Framework}
This section is mainly introductory. We collect the necessary
definitions and describe the main steps in the calculation
of inclusive decay widths to next--to--leading order.
\setcounter{equation}{0}
\subsection{The Effective Lagrangian}

Let us consider a Standard Model nonleptonic weak process
induced by charged currents. In the limit of infinite W boson mass
$m_W$ and
to leading order in the weak coupling, the charm--changing
nonleptonic decays of beautiful particles can be described by the
effective Lagrangian
\begin{eqnarray}\label{eq:effL}
{\cal L}_{\mbox{\scriptsize eff}}^{\Delta C =1} & = &
-\frac{G_F}{\sqrt{2}}\, V_{ud}^*V_{cb}^{\hphantom{*}}\left\{C_1(\mu)
{\cal O}_1(\mu)+ C_2(\mu) {\cal O}_2(\mu)\right\}\nonumber\\
& & {}-\frac{G_F}{\sqrt{2}}\, V_{us}^*V_{cb}^{\hphantom{*}}\left\{C_1(\mu)
{\cal O}'_1(\mu)+ C_2(\mu) {\cal O}'_2(\mu)\right\}.
\end{eqnarray}
Here ${\cal O}_2$ is the colour--singlet operator
\begin{equation}
{\cal O}_2 = (\bar du)_{V-A}\, (\bar c b)_{V-A}
\end{equation}
appearing in the tree--level Lagrangian, whereas the non--singlet
operator ($\alpha$ and $\beta$ are colour--indices)
\begin{equation}
{\cal O}_1 = (\bar d_\alpha u_\beta)_{V-A}\, (\bar c_\beta
b_\alpha)_{V-A}
\end{equation}
is generated by the strong interaction. The operators ${\cal O}'_i$
are obtained from the ${\cal O}_i$ by the replacement $\bar d\to \bar s$.
The Wilson--coefficients
$C_i(\mu)$ can be calculated at the scale $\mu = m_W$ and then evolved
down to scales $\mu\ll m_W$ by solving the corresponding
renormalization group equations. Since the operators ${\cal O}_1$ and
${\cal O}_2$ mix under renormalization, it is convenient to introduce
the operators ${\cal O}_{\pm} = ({\cal O}_2\pm {\cal O}_1)/2$ which
renormalize
multiplicatively.\footnote{Hereafter we imply that the
renormalization scheme obeys
 Fierz--symmetry. Within dimensional regularization this
requires certain restrictions on the particular choice of the so--called
evanescent operators, see \cite{BW90,JP94,HN94}.}
 To two--loop accuracy, the
corresponding Wilson--coefficients are given by \cite{ACMP81,BW90}
\begin{eqnarray}
C_\pm(\mu) & = & C_2(\mu)\pm C_1(\mu)\nonumber\\
& = & L_\pm(\mu)\!
\left\{1+\frac{\alpha_s(m_W)-\alpha_s(\mu)}{4\pi}
\frac{\gamma_\pm^{(0)}}{2\beta_0}\left(
\frac{\gamma_\pm^{(1)}}{\gamma_\pm^{(0)}}-\frac{\beta_1}{\beta_0}\right)
+\frac{\alpha_s(m_W)}{4\pi} B_\pm\right\}.\label{eq:wilsoncoeff}
\end{eqnarray}
Here
\begin{eqnarray}
L_\pm(\mu)  & =  & \left[ \frac{\alpha_s(m_W)}{\alpha_s(\mu)}
\right]^{\gamma_\pm^{(0)}/(2\beta_0)}
\end{eqnarray}
is the solution of the renormalization group equation for $C_\pm$ to
leading logarithmic order, whereas the term proportional to
$\alpha_s(m_W)-\alpha_s(\mu)$ takes into account the renormalization
group improvement to next--to--leading order. The $\gamma_\pm^{(i)}$
are the coefficients of the anomalous dimensions of the operators
${\cal O}_\pm$ and ${\cal O}'_\pm$,
\begin{equation}\gamma_\pm = \gamma_\pm^{(0)}\,\frac{\alpha_s}{4\pi} +
\gamma_\pm^{(1)}\left(\frac{\alpha_s}{4\pi}\right)^2 + {\cal O}(\alpha_s^3),
\end{equation}
with \cite{BW90}
\begin{equation}
\gamma_+^{(0)} = 4,\ \gamma_-^{(0)} = -8,\ \gamma_+^{(1)} = -7 +
\frac{4}{9}\, n_f,\ \gamma_-^{(1)} = -14 - \frac{8}{9}\, n_f
\end{equation}
in na\"{\i}ve dimensional regularization with anticommuting
$\gamma_5$ and $n_f$ running flavours (five in our case). The
$\beta_i$ are the first two
 coefficients of the QCD $\beta$--function,
\begin{eqnarray}
\beta &= & -g_s\left\{ \beta_0\,\frac{\alpha_s}{4\pi} + \beta_1\,
\left(\frac{\alpha_s}{4\pi}\right)^2 +
{\cal O}(\alpha_s^3)\right\},\nonumber\\
\beta_0 & = & 11-\frac{2}{3}\,n_f, \quad \beta_1 = 102 -
\frac{38}{3}\,n_f.
\end{eqnarray}
Finally,
the $B_\pm$ are the matching coefficients ensuring that up to
terms of order $\alpha_s^2(m_W)$, matrix elements of the effective
Lagrangian calculated at the scale $\mu=m_W$ equal the corresponding
matrix elements calculated with the full standard model Lagrangian.
Fierz--symmetry implies
\begin{equation}
\label{eq:Bpm}
         B_\pm = \pm B\, \frac{N_c \mp 1}{2N_c}\,,
\end{equation}
where $N_c=3$ is the number of colours. In the NDR scheme one finds
\cite{BW90}
\begin{equation}
     B=11.
\label{eq:B}
\end{equation}
Note that both the matching coefficients and the two--loop
anomalous dimensions depend on the renormalization scheme.
It proves, however, feasible to combine them into the
scheme--independent quantity \cite{ACMP81,GB93,BW90}
\begin{eqnarray}
R_\pm  & = & B_\pm + \frac{\gamma_\pm^{(0)}}{2\beta_0}\left(
\frac{\gamma_\pm^{(1)}}{\gamma_\pm^{(0)}}-\frac{\beta_1}{\beta_0}\right),
\nonumber\\
R_+ & = & \frac{10863 -1278 n_f+ 80 n_f^2}{6(33-2 n_f)^2}, \quad
R_- = -\frac{15021-1530n_f+80n_f^2}{3(33-2n_f)^2},
\end{eqnarray}
in terms of which the Wilson--coefficients read
\begin{equation}
C_\pm(\mu) =
L_\pm(\mu)\left\{1+\frac{\alpha_s(m_W)-\alpha_s(\mu)}{4\pi}
\,R_\pm+\frac{\alpha_s(\mu)}{4\pi}\, B_\pm\right\}.
\end{equation}
In this form, all the remaining scheme--dependence is absorbed
in $B_{\pm}$ and is to be cancelled by the scheme--dependence
of matrix elements of the corresponding operators.

\subsection{ The Inclusive Decay Rate}

The total inclusive decay rate of a
B meson into hadrons $X_c$, corresponding to the elementary subprocess
$b\to c \bar u d + c \bar u s$, can be expressed as the imaginary part of
a transition operator:
\begin{eqnarray}
\Gamma(B\to X_c) & = & \frac{\pi}{m_B}\sum\limits_{n}\!\int\!
\prod\limits_{i=1}^{n}\left[\frac{d^3\!p_i}{(2\pi)^32E_i}\right]
(2\pi)^3 \, \delta^4(p_B-\sum_i^n\,p_i)\,\left| \langle\,n\,|\,
{\cal L}_{\mbox{\scriptsize eff}}^{\Delta C=1}\,|\,B\,\rangle\right|^2
\makebox[0.6cm]{ }\nonumber\\
& = & \frac{1}{m_B}\,\mbox{Im}\,i\!\!\int\!\!d^4\!x\,\langle\, B\,|\,T\,
[{\cal L}_{\mbox{\scriptsize eff}}^{\Delta C=1}]^\dagger(x)\,
{\cal L}_{\mbox{\scriptsize eff}}^{\Delta C=1}(0)\,|\,B\,\rangle\,
\label{eq:scattampl}
\end{eqnarray}
where the sum runs over all possible final states.
Starting from this expression, it can be shown that in the limit of
infinite b quark mass the B meson
decay rate coincides with the decay rate of a free b quark, which can
be calculated perturbatively. Recently, much attention has been paid
to understand the structure of nonperturbative
corrections to that simple picture which are suppressed by
powers of the b quark mass \cite{general,nonlep}. Yet in this paper we
mainly
address the question of how to calculate radiative corrections to the
quark level decay and thus we proceed in a purely perturbative framework.

Summing over the squares of the relevant elements of the
CKM matrix and making use
of its approximate unitarity in the first two generations,
the hadronic decay width is equal to the width of the semileptonic decay
$b\to ce\bar\nu$,
\begin{equation}\label{eq:gamma0}
\Gamma_0 =
\frac{G_F^2\,m_b^5}{192\pi^3}\,|V_{cb}|^2\,f_1(m_c^2/m_b^2),
\end{equation}
multiplied by a colour factor 3 and up to radiative corrections. Here we
have neglected the light quark masses and the strange quark mass and
introduced the notation $f_1(a)$ for the tree-level phase--space factor:
\begin{equation}\label{eq:defPS}
f_1(a) = 1-8a+8a^3-a^4-12a^2\,\ln a.
\end{equation}
By the quark masses $m_b$ and $m_c$ we hereafter understand
 the pole masses.

\begin{figure}[h]
\centerline{
\epsfysize=0.7\textheight
\epsfbox{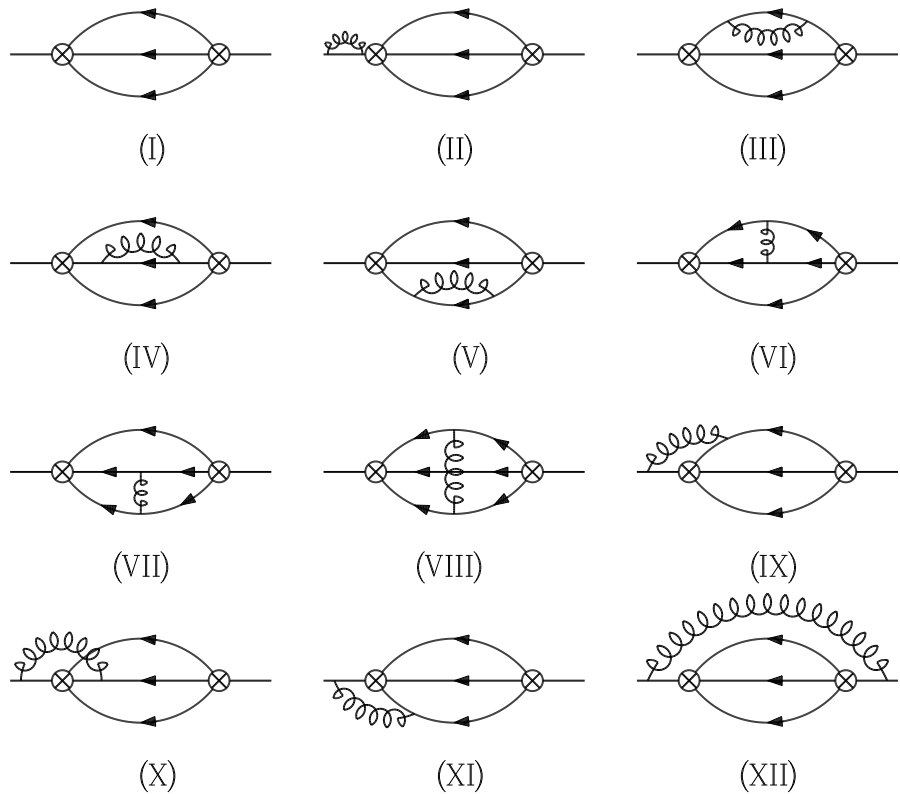}
}
\vspace{0.5in}
\caption[]{The diagrams contributing to the forward--scattering
amplitude Eq.\ (\protect{\ref{eq:FSA}}) up to order $\alpha_s$. The
crossed circles denote insertions of the operator
${\cal O}$. Of the three internal quark lines, the upper one denotes
the c quark, the lower one the d quark, and the middle one the u
antiquark.}\label{fig:1}

\end{figure}

The expression for the inclusive decay width becomes more complicated
if one turns on the
renormalization group running of the Wilson--coefficients in the
effective Lagrangian and calculates radiative corrections.
To next--to--leading order accuracy one
has to take into account the set of diagrams shown in
Fig.~\ref{fig:1}. For practical calculations it proves convenient
to split off the colour algebra
and trivial multiplicative factors. Then the operators
${\cal O}_1$ and ${\cal O}_2$ become identical and the four--fermion
vertices in Fig.~\ref{fig:1} can be interpreted as
insertions of ${\cal O} =(\bar d u)_{V-A} (\bar c b)_{V-A}$ or $(\bar
s u)_{V-A} (\bar c b)_{V-A}$ where {\em
no} summation over colour is implied. For definiteness,
the diagrams in Fig.\ \ref{fig:1}
represent the corresponding contribution of
\begin{equation}\label{eq:FSA}
 i\!\!\int\!\! d^4\! x\,\langle b|T[{\cal O}^\dagger(x){\cal O}(0)]
 |b\rangle_{\mbox{\scriptsize spin-averaged}}
\end{equation}
to the forward--scattering amplitude for an on--shell b quark.
The colour--factors of each diagram are given in Tab.~\ref{tab:1}.

Taking everything together, we can express the decay rate for $b\to
c\bar ud+ c\bar u s$ as
\begin{eqnarray}
\Gamma(b\to c\bar ud+ c\bar u s) & = &
\Gamma_0 \left[ 2L_+^2(\mu)+L_-^2(\mu) +
\frac{\alpha_s(m_W)-\alpha_s(\mu)}{2\pi} \left\{2L_+^2(\mu) R_++L_-^2(\mu)
R_-\right\}\right. \nonumber\\
& & {} + \frac{\alpha_s(\mu)}{2\pi}\left\{ 2L_+^2(\mu)\,B_+ +
L_-^2(\mu)\,B_-\right\}\nonumber\\
& & {} + \frac{3}{4}\,\left\{L_+(\mu)+L_-(\mu)\right\}^2 \,
\frac{2}{3}\,\frac{\alpha_s(\mu)}{\pi}\,\left\{ G_a + G_b \right\}
\nonumber\\ && {} +
\frac{3}{4}\,\left\{L_+(\mu)-L_-(\mu)\right\}^2\,\frac{2}{3}\,
\frac{\alpha_s(\mu)}{\pi}\,\left\{G_c + G_d\right\}\nonumber\\
& & \left.{}+\frac{1}{2}\,\left\{L_+^2(\mu)-L_-^2(\mu)\right\}\,
\frac{2}{3}
\,\frac{\alpha_s(\mu)}{\pi}\, \left\{G_a + G_b + G_e\right\}\right]
\label{eq:GammaFQD}
\end{eqnarray}
where $G_a, G_b, G_c, G_d$ and $G_e$ are functions of the ratio
$m_c/m_b$, and are obtained by adding the contributions
of particular diagrams, which we label by Roman numbers:
\begin{eqnarray}
\frac{\alpha_s}{\pi}\, G_a & =& K\,
  \mbox{Im (II + II}^\dagger\mbox{\ + III + IX +
IX}^\dagger\mbox{\ + XII),} \nonumber\\
\frac{\alpha_s}{\pi}\, G_b & = & K\,\mbox{Im (IV + V + VII),}\nonumber\\
\frac{\alpha_s}{\pi}\, G_c & = & K\, \mbox{Im (II + II}^\dagger
\mbox{\ + V + XI + XI}^\dagger \mbox{\ + XII),} \nonumber\\
\frac{\alpha_s}{\pi}\, G_d & = & K\,\mbox{Im (III + IV + VI),}\nonumber\\
\frac{\alpha_s}{\pi}\, G_e & = & K\,\mbox{Im (VI + VIII + X + X}^\dagger
\mbox{\ + XI + XI}^\dagger\mbox{).}\label{eq:defG}
\end{eqnarray}
Here
\begin{equation}\label{eq:defK}
   K = \frac{192 \pi^3}{m_b^6 f_1(m_c^2/m_b^2)}.
\end{equation}
Note that the colour--factors are split off
and shown explicitly in (\ref{eq:GammaFQD}).

It is easy to verify that $G_a$ and
$G_b$ are manifestly gauge--invariant and do not depend on the scheme
used for $\gamma_5$. The same is true for $G_c$ and $G_d$ provided the
renormalization scheme preserves Fierz--symmetry
 (cf.\ the next subsection). Taking into account the
relation (\ref{eq:Bpm}), the contribution of matching
coefficients in the second line in (\ref{eq:GammaFQD}) can equivalently
be rewritten replacing $G_e$ in the last line by $G_e+B$, which is again
a scheme--independent quantity.

The reason for splitting the full radiative correction to
the nonleptonic inclusive width into several pieces is that some of
these expressions also appear in the semileptonic decay widths.
For example, the $b\to c e \bar\nu $
decay rate can be written in terms of the function $G_a$ as
\begin{equation}\label{eq:defSLrate}
\Gamma(b\to c e\bar\nu) = \Gamma_0 \left\{ 1 +
\frac{2}{3}\,\frac{\alpha_s(\mu)}{\pi}\, G_a \right\}
\end{equation}
and, owing to Fierz--identities,
the function $G_c$ determines the radiative corrections to the decay
$b\to u\tau\bar\nu$,
\begin{equation}\label{eq:defSLlrate}
\Gamma(b\to u \tau\bar\nu) =\Gamma_0 \left\{ 1 +
\frac{2}{3}\,\frac{\alpha_s(\mu)}{\pi}\, G_c \right\},
\end{equation}
with the obvious replacements $V_{cb}\to V_{ub}$ and $a =
(m_c/m_b)^2 \to (m_\tau/m_b)^2$ in $\Gamma_0$ and $G_c$.
Since the radiative correction to the semileptonic decay
$b\to c e\bar\nu$ is known analytically for arbitrary c quark mass,
we can incorporate the answer given in
\cite{Nir89} directly in the expression for the
nonleptonic width and do not need to calculate the function
$G_a$ anew.

Finally, the calculation of  $G_b$ and $G_d$ is simplified by using
the known results for the radiative correction to
the vector correlation function with one massive quark:
\begin{eqnarray}
\Pi_{\mu\nu} &=& i\!\!\int\!\!
d^4\!x\,e^{iqx}\,\langle\,0\,|\,T[\bar{q}\gamma_\mu Q(x)\:
\bar{Q}\gamma_\nu q(0)]\,|\,0\,\rangle
\nonumber\\
&=& \int_{m_c^2}^\infty \frac{ds}{s-q^2}\left\{
q_\mu q_\nu \rho_1^V(s) + g_{\mu\nu}\rho_2^V(s)\right\} +
\mbox{constants}\,.
\label{eq:generalis}
\end{eqnarray}
The expressions for the spectral densities $\rho_i^V(s)$ can be
obtained from \cite{Gen90}; we have also calculated them directly
using Cutkosky rules (cf.\ App.~C).
Thus, essentially only $G_c$ and $G_e$ remain to be calculated, i.e.\
the diagrams II, V, VI, VIII, X, XI and XII.

It is worthwhile to note that the expressions for $G_a$ and $G_b$ are
available in form of one--dimensional parametric integrals for arbitrary
masses of all the three final state quarks from
\cite{HP}\footnote{We use the expressions given in the second of
Refs.~\cite{HP} and
imply that the factors $\ln\Delta/w$ appearing in Eq.~(3.35) of this
paper should be understood as $(\ln\Delta)/w$, as we have found by
comparison to the first of Refs.~\cite{HP}, and to the numerical results
given in \cite{HP}.}, corresponding to the upper and lower vertex
corrections,
respectively, in the notation of this paper. Answers for $G_c$ and
$G_d$ can be inferred from the expressions
given in \cite{HP} by Fierz--transformations. However, the calculation
of $G_e$ involves different Feynman diagrams than those calculated
 in \cite{HP} and is new.

\subsection{Calculational Tools}\label{subsection:2.3}

In this paper we use the well--known dispersion technique and
calculate the imaginary parts of the necessary diagrams
by explicitly summing the contributions of different discontinuities,
corresponding to real gluon emission and virtual corrections.
Infra--red (IR) divergences, which are bound to arise
due to the presence of massless particles in the final state, are
regulated by introducing non--zero gluon and light quark masses.
According to the famous Bloch--Nordsieck theorem, all IR
divergences cancel in the sum of all possible discontinuities
for a given diagram, so that in the final answer for the total
imaginary part one can put the regulator masses to zero.
In principle, one could escape the problem with the IR
divergences altogether calculating the matrix element in question
in Euclidian space, continuing analytically to Minkowki space
and taking the imaginary part. This technique is very powerful
for massless quarks, but, as we have found, is not very useful
when quark masses have to be taken into account. As for diagram V, we
prefer to express the subdiagram containing the self--energy by a
regularized dispersion
relation and only then compute the discontinuity.

We further use dimensional regularization and the
$\overline{\mbox{MS}}$ subtraction scheme to deal with
ultraviolet divergences, which give rise to the renormalization of
operators and quark masses.

\begin{figure}[h]
\centerline{
\epsfysize=0.25\textheight
\epsfbox{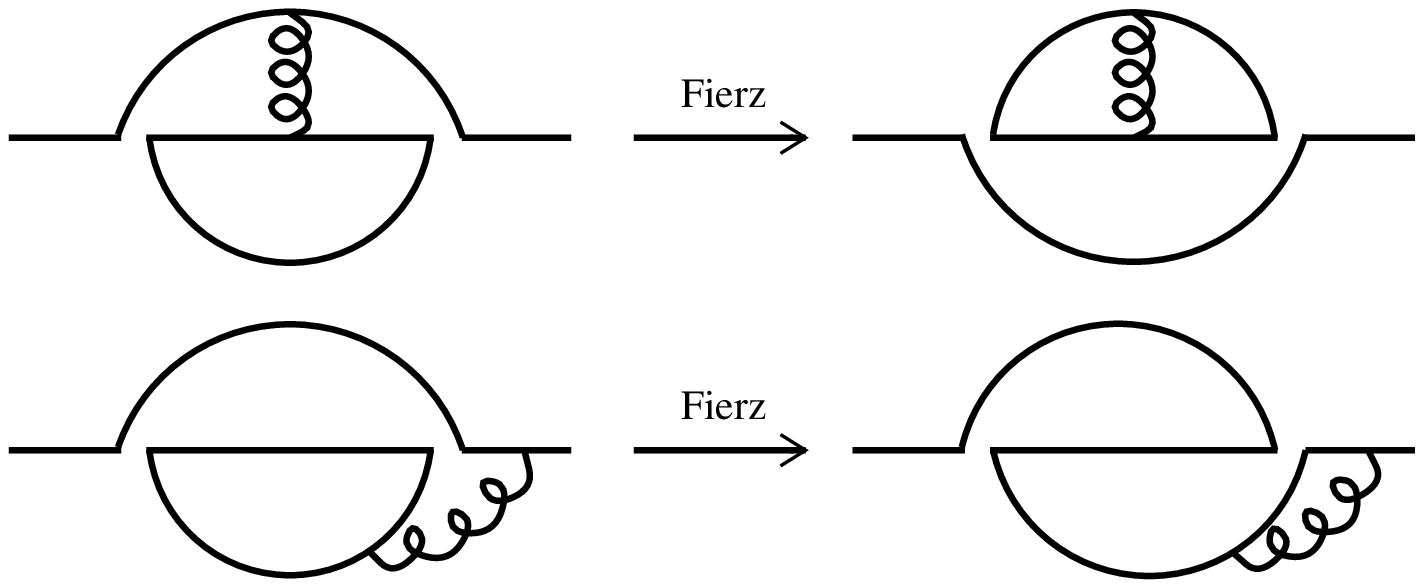}
}
\caption[]{Illustration of the effect of Fierz--transformations on the
 diagrams VI and XI. In contrast to the left--hand side diagrams, the
ones on the right--hand side are well defined in NDR.}\label{fig:2}

\vspace{0.5cm}

\centerline{
\epsfysize=0.35\textheight
\epsfbox{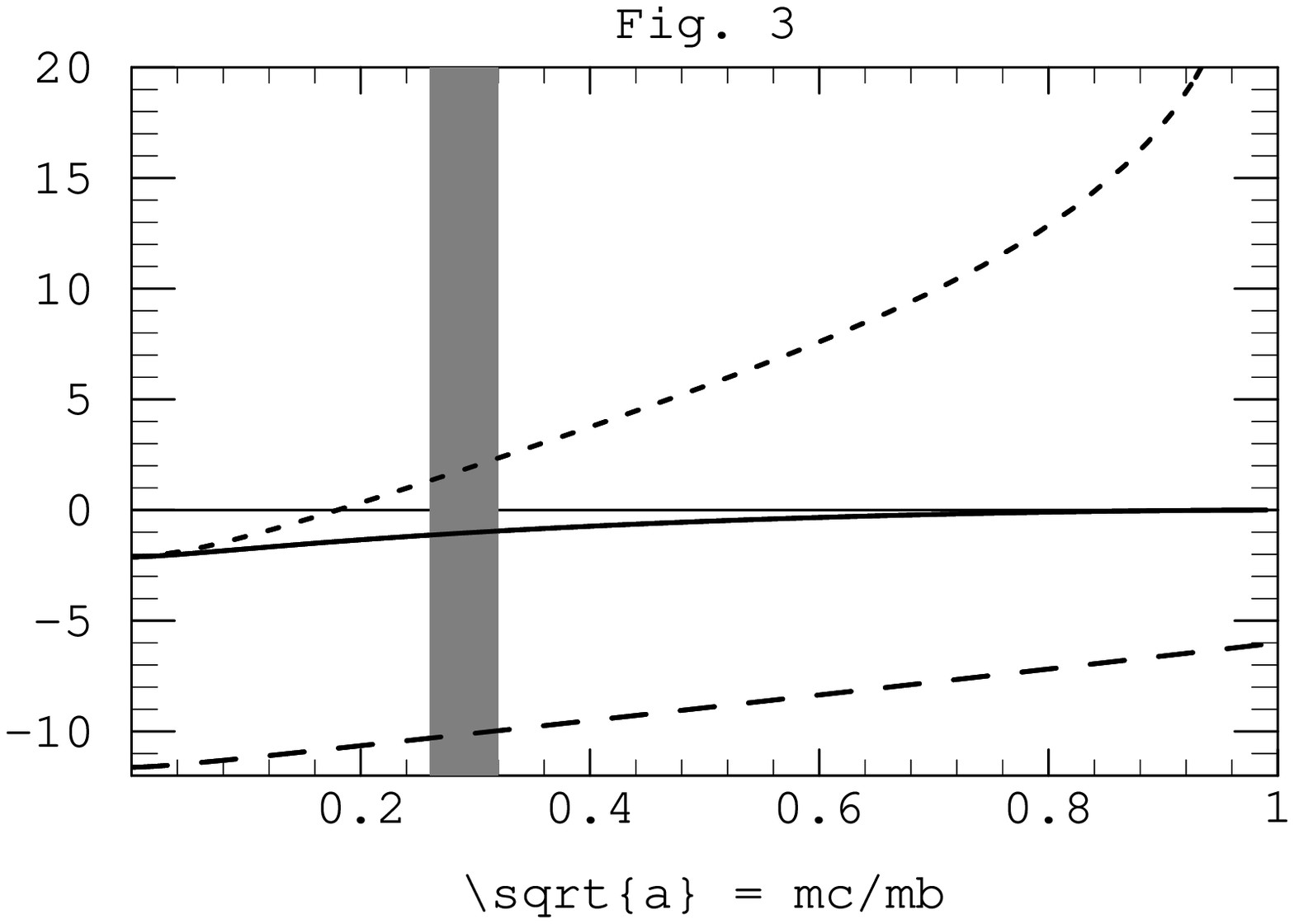}
}
\caption[]{Radiative corrections from the different combinations of
operator insertions as defined in Eq.\ (\protect{\ref{eq:cij}}) as
functions of $m_c/m_b$ for $\mu = m_b$: short dashes: $c_{11}$, long
dashes: $c_{12}$, solid line: $c_{22}$. The grey bar indicates the
range of realistic values of $m_c/m_b$. The numerical values are
tabulated in Tab.\ \protect{\ref{tab:2}}.}\label{fig:3}
\end{figure}

Due to the presence of
$\gamma_5$, the use of dimensional regularization
involves subtleties that have received a lot of attention over the last
years. We have found that for our purposes it
is sufficient to use the simplest prescription with
anticommuting $\gamma_5$ in  $D$ dimensions, NDR.
It is known that this scheme can yield algebraic
inconsistencies when the amplitude in question
contains closed fermion loops. Our method of calculation explicitly
avoids computing such diagrams by the appropriate use of
 Fierz--symmetry.
Indeed, by applying Fierz--transformations
one can always achieve that any diagram
contains no dangerous closed fermion
loops,\footnote{In fact, closed fermion loops with only
one external momentum and at most two open Lorentz--indices
 do not pose a problem, since
$\mbox{Tr}(\gamma_\mu\gamma_\nu \gamma_5)$ vanishes in any dimension.} as
illustrated in Fig.~\ref{fig:2}.
Subtleties are bound to arise, however,
 if the Fierz--symmetry is broken by
regularization. Thus, we use the results of Buras and Weisz \cite{BW90}
who have checked that in the NDR scheme
the renormalized operators ${\cal O}_1$ and ${\cal O}_2$
 have the same properties under Fierz--transformations
as the bare ones,
\begin{equation}
{\cal O}_2 \stackrel{\mbox{\scriptsize Fierz}}{\longrightarrow} {\cal
O}_1\, (c\leftrightarrow d),
\qquad {\cal O}_1
\stackrel{\mbox{\scriptsize Fierz}}{\longrightarrow} {\cal O}_2\, (c
\leftrightarrow d).
\end{equation}
Furthermore, with the particular choice of evanescent operators
made in \cite{BW90}
this property is valid {\em diagram by diagram}:
after subtraction of counterterms, renormalized Feynman
diagrams that can be related to each other by means of
Fierz--transformations in four dimensions coincide identically.
Note that we are only interested in the imaginary parts of the
multi--loop diagrams of Fig.~\ref{fig:1}, which are given by the
product of at most one--loop and tree--level amplitudes, followed by
phase--space integration. Thus, evaluating
the product with {\em renormalized} amplitudes, one can avoid
the contributions of evanescent operators altogether, provided
their choice is consistent with the calculation of
the one--loop matching coefficients
$B_\pm$ by a na\"{\i}ve projection onto four dimensions, see
\cite{BW90} for details.
 We  have checked that the alternative procedure for calculating the
whole three--loop diagram in dimension $D$, taking into account the
counterterms due to the evanescent operators as given in \cite{BW90},
yields the same result. Last but not least, we have found
that in the limit $m_c\to0$, our answers
 coincide with those obtained by Altarelli et
al., \cite{ACMP81}, using dimensional reduction, and by Buchalla,
\cite{GB93}, using the 't~Hooft--Veltman scheme, cf.\ App.~\ref{ss:mc}.
 More details about the calculation and some
useful intermediate formulas can be found in appendices A to C.

\section{Analytic results}
\setcounter{equation}{0}

In this section we give the explicit expressions for the
next--to--leading order corrections to the nonleptonic rate $b\to c
\bar u d$, the functions $G_x$ defined in Eq.\ (\ref{eq:defG}). The
answers for
all necessary single diagrams are given in App.~B. Throughout this
section, we use $a=(m_c/m_b)^2$; the factor
$K$ was defined in (\ref{eq:defK}) and
the phase--space function $f_1$ in (\ref{eq:defPS}). All results are
given in terms of pole masses.

As mentioned before, $G_a$ is proportional to
the semileptonic width $\Gamma(b\to c e \bar\nu)$. From the analytical
result given in \cite{Nir89} we obtain:
\begin{eqnarray}
{1\over K}{\alpha_s\over\pi}G_a & = & -{g_s^2 m_b^6\over9216\pi^5}
\left\{
-(1-a^2)(75-956 a+75 a^2)+4(1-a^2)\right.\nonumber\\
& & \times (17-64 a+17 a^2)\ln(1-a)
+4 a (60+270 a-4 a^2+17 a^3) \ln a\nonumber\\
& & {}-48(1+30 a^2+a^4) \ln a \;\ln(1-a) +12 a^2(36+a^2)\ln^2 a
\nonumber\\
& & {}-384 a^{3/2}(1+a)\left[\pi^2-2\ln{1-a^{1/2}\over1+a^{1/2}}
\ln a+ 4{\rm L}_2\left(-a^{1/2}\right)-
4{\rm L}_2\left(a^{1/2}\right)\right]
\nonumber\\
& & \left.{}- 12(1+16 a^2+a^4)(-\pi^2+6{\rm L}_2(a))\right\}.
\end{eqnarray}

We next calculate $G_b$ and $G_d$ by exploiting
the result for the vector correlation function  given in~\cite{Gen90}.
To get $G_d$, we first apply a Fierz--transformation to both
four--quark vertices of the diagrams III, IV and VI, cf.\ Fig.\
\ref{fig:2}. We then express the closed fermion loops by twice the
radiative
corrections to the vector correlation function\footnote{Note that the
vector correlation function equals the axial vector correlation
function in perturbative QCD.} $\Pi_{\mu\nu}$ defined
in Eq.\ (\ref{eq:generalis}) in their dispersion representation.
This effectively replaces the loop
by the propagator of a pseudoparticle, whose mass is just the invariant
mass of the original two fermions, multiplied by the weight functions
$\rho_i^{(2)}$ given in App.~C and followed by integration over
phase--space. By cutting the two remaining propagator lines we obtain
 the sum of imaginary parts of the diagrams ${\rm III} + {\rm IV} +
{\rm VI}$:
\begin{equation}\label{eq:anarel}
{1\over K}{\alpha_s\over \pi}G_d=
{1\over4\pi m_b^2}\int_{m_c^2}^{m_b^2} ds\;(m_b^2-s)^2
\left[m_b^2\rho^{(2)}_1(s)-2 \rho^{(2)}_2(s) \right].
\end{equation}
The integral can easily be done with the formulas given in App.~A. The
final result is:
\begin{eqnarray}
{1\over K}{\alpha_s\over \pi}G_d&=&
{g_s^2 m_b^6\over 9216\pi^5}\left\{(1-a)(18-476a-539a^2+75a^3)\right.
\nonumber\\& &{}+4a^2(36+8a-a^2)\left(\pi^2-3\ln^2 a \right)\nonumber\\
& &{}-2(1-a^2)\left[31-320a+31a^2-12(1-8a+a^2)\ln a\right]\ln(1-a)
\nonumber\\
& &{}-2a(132+90 a-308a^2+31 a^3)\ln a\nonumber\\
& &{}+24(2-16a-36a^2+8a^3-a^4+12 a^2\ln a ){\rm L}_2(a)\nonumber\\
& &\left.{}+ 864 a^2 [\zeta(3)-{\rm L}_3(a)]\right\}.
\end{eqnarray}
For $G_b$, the fermions in the loop are massless, so that
\begin{eqnarray}
{1\over K}{\alpha_s\over \pi}G_b&=&{1\over 4\pi m_b^2}
\int_0^{(m_b-m_c)^2} ds
\left\{
\left[(m_b^2-m_c^2)^2-(m_b^2+m_c^2) s
\right] \rho^{(2)}_1(s)\Big|_{\rm massless}\right.\nonumber\\
& &\left.{}-
2(m_b^2+m_c^2-s) \rho^{(2)}_2(s)\Big|_{\rm massless}\right\}
 w(s,m_b^2,m_c^2)\label{eq:Gb}\\
& = & {g_s^2 m_b^6\over 512\pi^5}f_1(a),\nonumber
\end{eqnarray}
where $w$ is defined as
\begin{equation}\label{eq:defw}
w(x,y,z) = (x^2+y^2+z^2-2xy-2xz-2yz)^{1/2}.
\end{equation}
Since $K$ is proportional to $1/f_1(a)$, the above result simply means
that $G_b$ is a constant independent of $a$.

Summing up the diagrams II, II${}^\dagger$, V, XI, XI${}^\dagger$ and
XII one gets
\begin{eqnarray}
{1\over K}{\alpha_s\over \pi}G_c&=&
{g_s^2 m_b^6\over 9216\pi^5}
\left\{
(1-a)(75-539 a-476 a^2+18 a^3)\right.\nonumber\\
& & {}-4(3-24a-36a^2+16a^3-2a^4)\pi^2\nonumber\\
& & {}-2(1-a^2)\left[31-320a+31a^2+12(1-8a+a^2)\ln a
\right]\ln(1-a)\nonumber\\
& & {}-2a(12+90a-188a^2+31a^3-48a\,\pi^2)\ln a\nonumber\\
& & {}-24(1-8a+36a^2+16a^3-2a^4+12a^2\ln a ){\rm L}_2(a)\nonumber\\
& &\left. {}+864 a^2[{\rm L}_3(a)-\zeta(3)]\right\}.
\end{eqnarray}

Finally, summing up the diagrams VI, VIII, X, X$^\dagger$, XI,
XI$^\dagger$, we obtain the scheme--independent quantity $G_e+B$:
\begin{eqnarray}
{1\over K}{\alpha_s\over \pi}\,(G_e+B)&=&{g_s^2 m_b^6\over 1536\pi^5}
\left\{-(1-a^2)(41-488a+41a^2)-16a(1+3a+a^2)[\pi^2-6{\rm L}_2(a)]
\right.\nonumber\\
& &{}-8(1-a^2)(1+28a+a^2)\ln(1-a)+4a^2(120-32a-5a^2)\ln a\nonumber\\
& &\left. {}+ 48a^3\ln^2 a+2(6\ln(m_b^2/\mu^2) + 11) f_1(a)
\right\}.\makebox[0.6cm]{ }\label{eq:3.7}
\end{eqnarray}
Note that $G_e$ contains a logarithm of $\mu$, namely
 $G_e = 6\ln(m_b^2/\mu^2)+\dots$, which just cancels the remaining
$\mu$ dependence of the leading order term in Eq.\ (\ref{eq:GammaFQD}),
 as shown by an explicit
expansion around $\mu= m_b$:
\begin{equation}
2L_+^2(\mu) + L_-^2(\mu) = 2L_+^2(m_b) + L_-^2(m_b) -
\frac{1}{2}\,\left\{L_+^2(m_b) - L_-^2(m_b) \right\}\frac{2}{3}\,
\frac{\alpha_s(m_b)}{\pi}\:6\,\ln\,\frac{m_b^2}{\mu^2}.
\end{equation}

This completes the list of entries required to evaluate
 the radiative corrections to the nonleptonic decay $b\to c\bar u d$.
We have checked that our expressions for $G_c$ and $G_d$ agree numerically
to the results given in \cite{HP}. As mentioned above, the answer for
$G_e$ cannot be inferred from this paper.

For practical calculations, it may be useful to give approximations
to the above formulas for small $a$:
\begin{eqnarray}
(G_a+G_b)f_1(a)&=&{31\over4}-\pi^2-a[80+24\ln a]+32a^{3/2}\,\pi^2
-a^2 [273+16\pi^2-18\ln a\nonumber\\
& & {}+36\ln^2 a]+32 a^{5/2}\,\pi^2-{8\over9}\,a^3\,
\left[118- 57\ln a\right]+{\cal O}\left(a^{7/2}\right),\\
(G_c+G_d)f_1(a)&=&{31\over4}-\pi^2-8a[10-\pi^2+3\ln a]
-a^2 [117-24\pi^2\nonumber\\
& & {}+(30-8\pi^2)\ln a+36\ln^2 a]-{4\over3}\,a^3\,[79+2\pi^2\nonumber\\
& & {}- 62\ln a+6\ln^2 a]+{\cal O}\left(a^4\right),\\
(G_a+G_b+G_e+B)f_1(a)&=&\left(6\ln{m_b^2\over\mu^2}+11\right)f_1(a)-
{51\over4}-\pi^2+8a[21-\pi^2-3\ln a ]\nonumber\\
& &{}+32 a^{3/2}\,\pi^2 - a^2 [111+40\pi^2-258\ln a+36\ln^2 a]
\nonumber\\
& &{}+32 a^{5/2}\,\pi^2-{4\over9}\,a^3\,[
305+18\pi^2  + 30\ln a-54\ln^2 a]\nonumber\\
& &{}+{\cal O}\left(a^{7/2}\right).
\end{eqnarray}
These approximations are valid for $a<0.15$, which covers a
realistic range of values $(m_c/m_b)^2$.

The corresponding approximate formula for
the semileptonic decay $b\to c e \bar\nu$ is given in~\cite{Nir89}. We repeat
it here for completeness, converting to our notations, cf.\
(\ref{eq:defSLrate}):
\begin{eqnarray}
G_a f_1(a) &=&{25\over4}-\pi^2-a[68+24\ln a ]+32 a^{3/2}\pi^2
-a^2[16\pi^2+273-36\ln a +36 \ln^2 a]\nonumber\\
& &{}+32 a^{5/2} \pi^2-a^3\left[{1052\over9}-{152\over3}\ln a \right]+
{\cal O}\left(a^{7/2}\right).
\end{eqnarray}
Similarly, we obtain for $b\to u\tau\bar\nu$, cf.\ (\ref{eq:defSLlrate}):
\begin{eqnarray}
 G_c f_1(a)&=&{25\over4}-\pi^2-8 a[6-\pi^2 ]
-2 a^2[15-6\pi^2+(15-4\pi^2)\ln a+36 \zeta(3)]\nonumber\\
& & {}+a^3\left[{20}-{16\pi^2\over3} \right]+{\cal O}\left(a^4\right),
\end{eqnarray}
where here $a$ has to be interpreted as $(m_\tau/m_b)^2$.

These expressions complete our formulas of radiative corrections
to b quark decays with one massive particle in the final state.

\section{Estimate of Finite Mass Corrections to $\Gamma(b\to c\bar c
s)$}
\setcounter{equation}{0}

The structure of the decay rate $\Gamma(b\to c\bar c s)$ is very
similar to $\Gamma(b\to c\bar u d)$, except for the additional
contributions of penguin diagrams. Since penguin operators enter the
effective Lagrangian only by mixing, their Wilson--coefficients are
small and thus penguin contributions can usually be neglected in
tree--level decays, cf.\ \cite{AP91}. Defining functions $H_x$ in a
completely analogous way to the $G_x$ that enter $\Gamma(b\to c\bar u
d)$, Eq.\ (\ref{eq:GammaFQD}), the decay rate of $b\to c\bar c s+c\bar
c d$ can be written as
\begin{eqnarray}
\lefteqn{\kern-1.8cm\Gamma(b\to c\bar cs + c\bar cd) =
\Gamma_0' \left[ 2L_+^2(\mu)+L_-^2(\mu) +
\frac{\alpha_s(m_W)-\alpha_s(\mu)}{2\pi} \left\{2L_+^2(\mu) R_++
L_-^2(\mu) R_-\right\}\right.} \nonumber\\
& & {} + \frac{\alpha_s(\mu)}{2\pi}\left\{ 2L_+^2(\mu)\,B_+ +
L_-^2(\mu)\,B_-\right\}\nonumber\\
& & {} + \frac{3}{4}\,\left\{L_+(\mu)+L_-(\mu)\right\}^2 \,
\frac{2}{3}\,\frac{\alpha_s(\mu)}{\pi}\,\left\{ H_a + H_b \right\}
\nonumber\\ && {} +
\frac{3}{4}\,\left\{L_+(\mu)-L_-(\mu)\right\}^2\,\frac{2}{3}\,
\frac{\alpha_s(\mu)}{\pi}\,\left\{H_c + H_d\right\}\nonumber\\
& & \left.{}+\frac{1}{2}\,\left\{L_+^2(\mu)-L_-^2(\mu)\right\}\,
\frac{2}{3}
\,\frac{\alpha_s(\mu)}{\pi}\, \left\{H_a + H_b + H_e\right\}\right]
+{\rm penguins}.\label{eq:gammaprime}
\end{eqnarray}
The tree--level decay rate $\Gamma_0'$ is defined as in Eq.\
(\ref{eq:gamma0}) except for the phase--space factor. Neglecting the
strange quark mass, $f_1(a)$ has to be replaced by
\begin{equation}\label{eq:f1aa}
f_1(a,a) =
{\sqrt{1 - 4\,a}}\,\left( 1 - 14\,a - 2\,{a^2} - 12\,{a^3} \right)  +
   24\,{a^2}\,\left( 1 - {a^2} \right) \,
  \ln \left({{1 + {\sqrt{1 - 4\,a}}}\over {1 - {\sqrt{1 - 4\,a}}}}\right)
\end{equation}
with $a = (m_c/m_b)^2$. With $m_s=0.2\,$GeV, (\ref{eq:f1aa}) is
accurate to better than 5\% for $a<0.35^2$. In the subsequent
estimate of $\Gamma(b\to c\bar{c} s)$, however, we take into account
a finite $m_s$ and use the three--particle phase--space factors
calculated in \cite{CPT82}.

$H_a$ parametrizes the corrections to the decay rate $b\to c \tau\bar\nu$
with the ``$\tau$'' and the c quark degenerate in mass, whereas $H_c$
now
describes the corrections to a decay  of type $b\to u \tau\bar\nu$, where
the light quark $u$ is still massless, but the {\em leptons} have
equal {\em finite} masses.

The calculation of $H_b$ requires only expressions already
obtained in this paper. From Eq.\ (\ref{eq:Gb}), where $G_b$ is
expressed as a single integral over spectral functions, we immediately
obtain:
\begin{eqnarray}
{1\over K'}{\alpha_s\over \pi}H_b&=&{1\over 4\pi m_b^2}
\int_{m_c^2}^{(m_b-m_c)^2} ds
\left\{
\left[(m_b^2-m_c^2)^2-( m_b^2+m_c^2) s
\right] \rho^{(2)}_1(s)\right.\nonumber\\
& &\left.{}-
2(m_b^2+m_c^2-s) \rho^{(2)}_2(s)\right\}
 w(s,m_b^2,m_c^2)\label{eq:Hb}
\end{eqnarray}
with $w$ defined in (\ref{eq:defw}) and $\rho_i^{(2)}$ given in App.~C.
$K'$ equals $K$, Eq.\ (\ref{eq:defK}), except for the obvious
replacement of phase--space factors, $f_1(a)\to f_1(a,a)$.
$H_b$ is extremely sensitive to the value of the c quark mass as
clearly visible from Tab.\ \ref{tab:3}; recall that the corresponding
quantity for $b\to c\bar u d$, $G_b$, was constant in $m_c$!
The numerical values given in Tab.~\ref{tab:3} agree with the
corresponding ones obtainable from \cite{HP}. In the table we also
give the values of $H_a$, $H_c$ and $H_d$ as functions of $m_c/m_b$
calculated from the formulas in \cite{HP}. $H_e$ cannot be obtained
from the work of Hokim and Pham, so for the present estimate we
insert $H_e \approx G_e$, Eq.\ (\ref{eq:3.7}), and estimate the error
introduced by this procedure by twice the difference between $G_e(a)$
and $G_e(0)$. From Tab.~\ref{tab:3} it is clear that both $H_a+H_b$
and $H_c+H_d$ are extremely sensitive to $m_c$. For the realistic
value $m_c/m_b\approx 0.3$, both $H_a+H_b$ and $H_c+H_d$ are positive
and by a factor four to five larger than the corresponding $G_x$. In
Tab.~\ref{tab:x} we also give the decay rate $\Gamma(b\to c\bar c s)$
including the error in $H_e$. For $m_c/m_b=0.3$, the finite c quark
mass effects in the radiative corrections increase the decay rate by 30\%.

\section{Numerical Results}
\setcounter{equation}{0}

Let us now turn to the numerical analysis of the results obtained in
Secs.~3 and 4. To do this, we first rewrite the decay width Eq.\
(\ref{eq:GammaFQD}) in a form that is more convenient in the present
context ($a = (m_c/m_b)^2$). Introducing scheme--independent
coefficients $c_{ij}$ by $c_{11} = G_c+G_d$, $c_{12} = G_a+G_b+G_e+B$
and $c_{22}=G_a+G_b$, we have
\begin{eqnarray}
\lefteqn{\Gamma(b\to c\bar ud+ c\bar u s) =
\Gamma_0 \left[ 2L_+^2(\mu)+L_-^2(\mu) +
\frac{\alpha_s(m_W)-\alpha_s(\mu)}{2\pi} \left\{2L_+^2(\mu) R_++L_-^2(\mu)
R_-\right\}\right.} \nonumber\\
& & {} + \frac{3}{4}\,\left\{L_+(\mu)+L_-(\mu)\right\}^2 \,
\frac{2}{3}\,\frac{\alpha_s(\mu)}{\pi}\,c_{22}(a)
+\frac{3}{4}\,\left\{L_+(\mu)-L_-(\mu)\right\}^2\,\frac{2}{3}\,
\frac{\alpha_s(\mu)}{\pi}\,c_{11}(a)\nonumber\\
& & \left. {}+\frac{1}{2}\,\left\{L_+^2(\mu)-L_-^2(\mu)\right\}\,
\frac{2}{3}\,\frac{\alpha_s(\mu)}{\pi}\,c_{12}(a)\right]
\label{eq:cij}\\
& \equiv & 3\,\Gamma_0\,\eta(\mu)\,J(a,\mu).\label{eq:defJ}
\end{eqnarray}
Here we use the same notations as \cite{AP91} with
\begin{equation}
\eta(\mu) = \frac{1}{3} \{2L_+^2(\mu)+L_-^2(\mu)\},
\end{equation}
whereas $J = 1+{\cal O}(\alpha_s)$ covers the next--to--leading order
terms.

In Fig.\ \ref{fig:3}, the coefficients $c_{11}$, $c_{12}(\mu=m_b)$
and $c_{22}$ are plotted as functions of $m_c/m_b$; we have also
tabulated the numerical values in Tab.\ \ref{tab:2}. A range of
realistic values of $m_c/m_b$, $0.26\leq m_c/m_b\leq 0.32$, is
indicated by a grey bar. The figure shows
clearly that the colour--favoured coefficient $c_{22}$ is not very
sensitive to the c quark mass. This is due to the fact that $G_b\equiv
3/2$ is constant in $m_c$, whereas the term $G_a$, which also
determines the semileptonic decay width, exhibits an only mild
dependence on
$m_c$. The coefficient $c_{12}$ increases more steeply in $m_c$, but
is suppressed in the rate by a small Wilson--coefficient. Both $c_{12}$
and $c_{22}$ are finite for $m_c\to m_b$. This is different for
$c_{11}$: this function diverges like $\sim \ln (m_b-m_c)$
as shown in Tab.\ \ref{tab:2}. Note, however,
that the decay rate still vanishes for $m_c=m_b$. For $c_{11}$ and
$c_{22}$ we found agreement with the integrals given in \cite{HP}.

\begin{figure}[h]
\makebox[1cm]{}
\vskip-0.6in
\centerline{
\epsfysize=0.55\textheight
\epsfbox{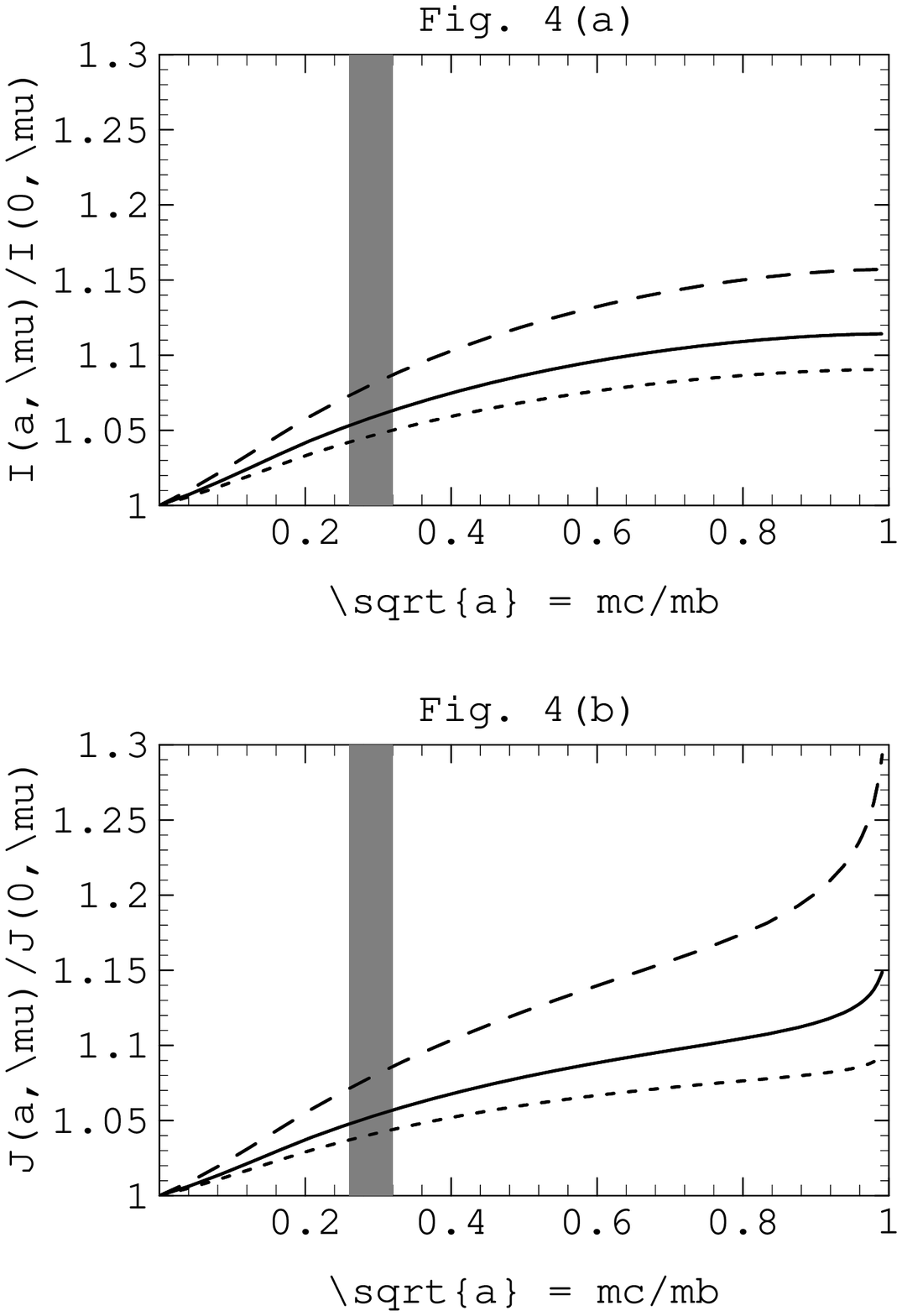}
}
\caption[]{(a) The next--to--leading order correction $I(a,\mu)$ to
$\Gamma(b\to c e \bar \nu)$, Eq.\ (\protect{\ref{eq:defI}}), as
function of $\sqrt{a} = m_c/m_b$, normalized to one at $m_c=0$, for
$\alpha_s(m_Z) = 0.117$ and three
different choices of the renormalization scale: solid line: $\mu =
m_b$, long dashes: $\mu = m_b/2$, short dashes: $\mu = 2 m_b$.
(b) The next--to--leading order correction $J(a,\mu)$ to
$\Gamma(b\to c \bar u d)$, Eq.\ (\protect{\ref{eq:defJ}}), as
function of $\sqrt{a} = m_c/m_b$, normalized to one at $m_c=0$ and
using the same parameters as in (a). The grey bar
indicates the range of realistic values of $m_c/m_b$.}\label{fig:4}

\vspace{10pt}

\centerline{
\epsfysize=0.25\textheight
\epsfbox{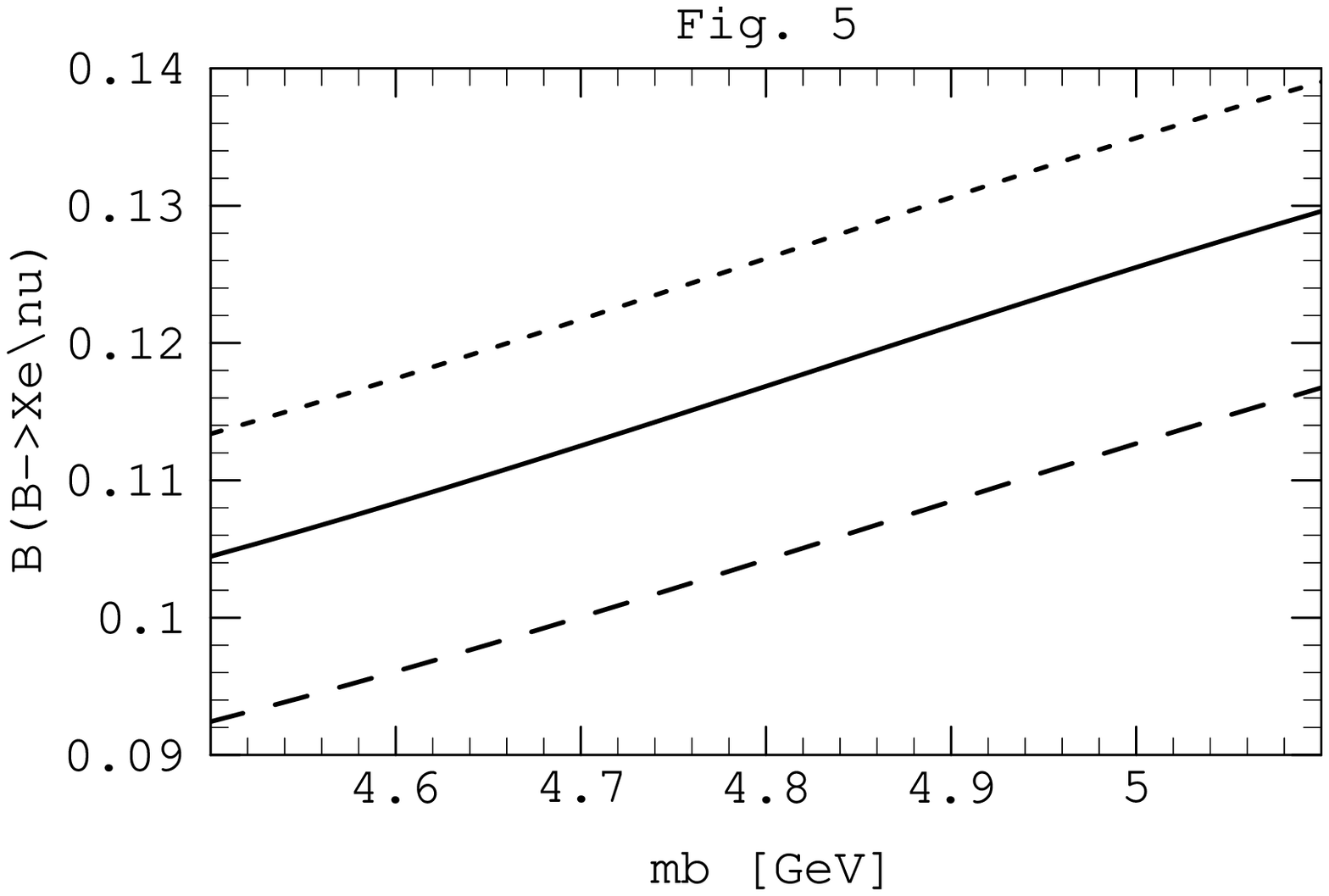}
}
\caption[]{The semileptonic branching ratio of the B meson including
non--perturbative corrections as a function of $m_b$
with $m_b-m_c$ fixed by HQET and $\alpha_s(m_Z) = 0.117$. The three
lines correspond to three different
choices of the renormalization scale: solid line: $\mu = m_b$, long
dashes: $\mu = m_b/2$, short dashes: $\mu = 2 m_b$.}\label{fig:5}
\end{figure}

We next investigate the behaviour of $\Gamma(b\to c\bar u d)$ as
compared to $\Gamma(b\to c e\bar\nu)$. To this purpose, we introduce
\begin{equation}\label{eq:defI}
I(a,\mu) = 1+\frac{2}{3}\,\frac{\alpha_s(\mu)}{\pi}\, G_a(a),
\end{equation}
so that $\Gamma(b\to c e\bar\nu) = \Gamma_0\,I(a,\mu)$.
In Fig.\ \ref{fig:4}(a) we plot $I(a,\mu)/I(0,\mu)$ as function
of $m_c/m_b$ for three different values of the renormalization scale
$\mu$. Fig.\ \ref{fig:4}(b) shows the ratio
$J(a,\mu)/J(0,\mu)$ as function of $m_c/m_b$. For $m_c/m_b\approx
0.3$, we find an increase of the nonleptonic decay rate $b\to c\bar u
d$ by (4--8)\%, depending on the renormalization scale. Comparing the
two plots, we find that in the range of realistic values of $m_c/m_b$,
which is emphasized by a grey bar, the enhancement by finite mass
corrections is approximately equal for semi-- and nonleptonic decays,
whereas for $m_c/m_b$ near one, the logarithmic divergence in $c_{11}$
clearly dominates.

Whereas up to now we have discussed only quantities that are defined
on a purely partonic level, one clearly is more interested in
predictions of {\em hadronic} decay rates and branching ratios. Though
in general this requires the knowledge of hadronic matrix elements,
matters simplify considerably for inclusive decays of heavy hadrons.
Without going into further details, we just mention that the theory
of inclusive decays of heavy hadrons has
experienced considerable progress over the last years when
another field of application of the
well--known operator product expansion (OPE) technique
as an expansion in inverse powers of the heavy quark mass
was opened, see \cite{general,nonlep}. The situation is even
more favourable than in other fields since the leading
term in the heavy quark expansion does not involve {\em any} hadronic
uncertainties, but simply coincides with the underlying free quark
decay process. Hadronic corrections enter only at second and higher
order in the expansion with a natural size of
$\sim 1\,{\rm GeV}^2/m_b^2\sim 5\%$.

To second order in the heavy quark expansion, these corrections can
be expressed in terms of two matrix elements,
\begin{eqnarray}
2m_B\lambda_1 & = & \langle\,B\,|\,\bar{b}_v (iD)^2
b_v\,|\,B\,\rangle, \nonumber\\
6m_B\lambda_2 & = & \langle\,B\,|\,\bar{b}_v \frac{g}{2}\,
\sigma_{\mu\nu} F^{\mu\nu}b_v\,|\,B\,\rangle,
\end{eqnarray}
where $b_v$ is defined as $b_v = e^{im_b v x}b(x)$, $b(x)$ being the
$b$ quark field in full QCD, $v_\mu$ is the four--velocity of
the B meson, $m_B$ its mass and $F^{\mu\nu}$ the
gluonic field--strength tensor.

Whereas $\lambda_2$ is directly related to the observable
spectrum of beautiful mesons,
\begin{equation}
\lambda_2 \approx \frac{1}{4}\,(m_{B^*}^2 - m_B^2)= 0.12\,{\rm GeV}^2,
\end{equation}
the quantity $\lambda_1$ is difficult to measure, cf.\
\cite{grozin}. Physically, $-\lambda_1/(2m_b)$ is
just the average kinetic energy of the b quark inside the meson.
At present, only a QCD sum rule estimate is available,
according to which $\lambda_1 \simeq -0.6$ GeV$^2$ \cite{BB94}.
 This result has been met with caution
(see, e.g.\ \cite{vir}), since it corresponds in fact to a surprisingly
large momentum of the b quark inside the meson of order (700--800)$\,$MeV.
However, in a recent series of papers, cf.\ \cite{grozin,SVlimit},
an upper bound on $\lambda_1$ has been derived, to wit
$\lambda_1 \leq -0.4\,{\rm GeV}^2$,
which appears to be in nice agreement with the QCD
sum rule prediction. In our analysis we thus conform to the value
$\lambda_1 =  -(0.6\pm 0.1)\,{\rm GeV}^2$.

Taking into account hadronic corrections, the
nonleptonic decay rate
of a B meson into a single charmed hadronic state
can be written as \cite{nonlep}:
\begin{eqnarray}
\Gamma(B\to X_c) & = & 3 \Gamma_0 \left[ \eta(\mu) J(a,\mu) \left\{ 1 +
\frac{\lambda_1}{2m_b^2} + \frac{3\lambda_2}{2m_b^2} - \frac{6
(1-a)^4}{f_1(a)} \frac{\lambda_2}{m_b^2}\right\}\right.\nonumber\\
& &\left.{} - \left\{L_+^2(\mu) -
L_-^2(\mu)\right\} \frac{4(1-a)^3}{f_1(a)} \frac{\lambda_2}{m_b^2} +
{\cal O}\left(\alpha_s^2,\,\frac{\alpha_s}{m_b^2},\,\frac{1}{m_b^3}\right)
\right].
\end{eqnarray}
For the semileptonic rate one obtains:
\begin{equation}
\Gamma(B\to X_c e \bar\nu) = \Gamma_0\left[I(a,\mu)\left\{ 1 +
\frac{\lambda_1}{2m_b^2} + \frac{3\lambda_2}{2m_b^2} - \frac{6
(1-a)^4}{f_1(a)} \frac{\lambda_2}{m_b^2}\right\}
 + {\cal O}\left( \alpha_s^2,\,\frac{\alpha_s}{m_b^2},\,\frac{1}{m_b^3}
\right) \right].
\end{equation}

In evaluating these decay rates, it is crucial to minimize
the number of independent parameters. In doing so, we
take advantage of the fact that the {\em difference} between
heavy quark masses is fixed in the framework of the heavy quark expansion:
\begin{equation}\label{eq:gaehn}
m_b - m_c = m_B-m_D + \frac{\lambda_1+3\lambda_2}{2}\,\left(
\frac{1}{m_b} -\frac{1}{m_c} \right) + {\cal O}
\left(\frac{1}{m^2}\right).
\end{equation}
The only quantity remaining to be fixed is then $m_b$ or $m_c$. We
prefer to take $m_b$ from spectroscopy and choose the most
conservative range\footnote{Since we give a broad range of values, we
are not concerned with the intrinsic uncertainty in the
definition of the pole
mass which is caused by renormalons and estimated to be of order
(50--200)$\,$MeV, \cite{renormalons}.}
\begin{equation}
4.5\,{\rm GeV}\leq m_b\leq 5.1\,{\rm GeV}.
\end{equation}
With these values we find:
\begin{equation}\label{eq:thefinalone}
\frac{\Gamma(B\to X_c)}{\Gamma(B\to X_c e \bar\nu)} =  4.0\pm 0.4.
\end{equation}
This result is nearly independent of the c quark mass; the error is
entirely due to the dependence on the renormalization scale, $m_b/2\leq
\mu \leq 2 m_b$. Note that it is mainly the mass--independent
$G_b$ that enters the radiative corrections, whereas $G_a$ cancels,
which explains the small sensitivity of (\ref{eq:thefinalone}) to
$m_c$.

Finally we calculate the semileptonic branching ratio as
\begin{equation}\label{eq:ultima}
B(B\to X e \bar\nu) = \frac{\Gamma(B\to X e \bar\nu)}{\sum_{\ell =
e,\,\mu,\,\tau}\Gamma(B\to X\ell \bar\nu_{\ell}) + \Gamma(B\to X_c) +
\Gamma(B\to X_{c\bar c})}.
\end{equation}
Previous analyses yielded $B(B\to Xe\bar\nu)>12.5\%$
\cite{AP91,BBSV941} which is considerably bigger than what is measured
experimentally, $B(B\to Xe\bar\nu) = (10.43\pm 0.24)\%$ \cite{PartData}.
Yet these studies were lacking
the finite quark mass effects on $J(a,\mu)$, the radiative
corrections to $\Gamma(b\to c\bar cs)$ and the corrections to
$\Gamma(b\to c\tau\bar\nu)$.
Taking them into account, we find $B(B\to
Xe\bar\nu)= (11.6\pm 1.8)\%$ where the error comes from the uncertainty
in the renormalization scale, in $\lambda_1$, in the quark mass and in
the value of $\alpha_s(m_Z)$, where we used $\alpha_s(m_Z) = (0.117\pm
0.007)$ \cite{bethke}. The functional dependence of the semileptonic
branching ratio on $m_b$ with $m_c$ fixed by (\ref{eq:gaehn}) is
also plotted in Fig.\ \ref{fig:5}. Note that in $\Gamma(B\to X_{c\bar
c})$ in Eq.\ (\ref{eq:ultima}) we use the full tree--level
phase--space factor calculated in \cite{CPT82} including a strange
quark mass $m_s = 0.2\,$GeV. With that value of $m_s$, the rate is
decreased by typically 5\%, which is a much smaller effect than taking
into account the c quark mass in the radiative corrections.

In Tab.~\ref{tab:extra} we compare
the semileptonic branching ratios obtained in \cite{AP91,BBSV941} with
ours, using the same input parameters. It turns out that, although the
introduction of the heavy quark expansion by Bigi et al.\ leads to a
reduction of the branching ratio by 0.3\% compared to the quark level
analysis of Altarelli and Petrarca, the finite c
quark mass effects in the radiative corrections to the nonleptonic
widths are by more than a factor three bigger and yield an additional 1.0\%
reduction. This fact shows the paramount importance of {\em perturbative}
corrections in the heavy quark expansion.

Yet, we consider this analysis as a rather preliminary one. To make
more definite predictions, one has to reduce the scheme--dependence.
We plan to come back to
this and related questions in a separate publication \cite{B^3G}.

\section{Summary and Conclusions}

In this paper we have calculated the radiative corrections to
the free quark decay $b\to c\bar u d$ as the imaginary part of the
relevant forward--scattering amplitude including the full dependence
on the c quark mass. We have performed the calculation
in na\"{\i}ve dimensional regularization (NDR) with anticommuting
$\gamma_5$. The key--observation that allowed us to employ this scheme
despite of its known deficiencies was that diagrams that are ambiguous
in NDR can be related to well--defined ones by means of
Fierz--transformations.
In the limit $m_c\to 0$, our results agree with known expressions
obtained in other schemes. As far as comparison is possible, our
results also agree with those of \cite{HP}.

For realistic values of the b and the c quark mass, we find a moderate
increase of about (4-8)\% of the decay rate $\Gamma(b\to c\bar u d)$
with respect to the limit of vanishing c quark mass, depending on the
renormalization scale.

We also have estimated the increase of $\Gamma(b\to c\bar c s)$ by
finite mass effects to be about 30\%. The
calculation of the full radiative corrections is under study.

In the framework of the heavy quark expansion of inclusive decays of
heavy hadrons, the knowledge of finite quark mass corrections to
nonleptonic decays is crucial for an improvement of the theoretical
prediction of various measurable quantities like $\Gamma(B\to X_c)/
\Gamma(B\to X_c e\bar\nu)$ and $B(B\to Xe\bar\nu)$. We have shown that the
ratio of non-- to semileptonic decays of B mesons to single charmed
final states remains nearly unaffected by finite mass effects, whereas
the semileptonic branching ratio is reduced by 1.0\% compared with
a recent investigation by Bigi et al., \cite{BBSV941}. The finite mass
terms in the radiative corrections are thus of the same importance as
the hadronic corrections introduced in \cite{general,nonlep} and
severely reduce
the scope of possible new physics in nonleptonic B decays. Yet, the
analysis of
the semileptonic branching ratio involves delicate points like the
choice of quark masses and estimates of higher order
radiative corrections; we will address  these points
 in a separate paper \cite{B^3G}.

\section*{Acknowledgements}

P.~B. and V.~B. would like to thank B.~Stech for bringing the problem
to their attention.
P.~B. is grateful to  A.~Buras, S.~Herrlich, M.~Lautenbacher,
M.~Misiak and U.~Nierste for discussions about evanescent operators and
related topics. V.~B. gratefully acknowledges correspondence with
X.Y.~Pham concerning Ref.~\cite{HP}.
P.~B. and V.~B. especially thank the Grup de F\'{\i}sica
Te\`{o}rica of the Universitat Aut\`{o}noma de Barcelona for the warm
hospitality extended to them during their stay there where part of this
work has been done.  E.~B. would like to thank BNL's Theory Group for
its hospitality and acknowledges the financial support of the CYCIT,
project No.\ AEN93-0520. P.~G. acknowledges
gratefully a grant from the Generalitat de Catalunya.

\appendix
\renewcommand{\theequation}{\Alph{section}.\arabic{equation}}

\section*{Appendix A}
\setcounter{equation}{0}
\addtocounter{section}{1}

In this appendix we collect some intermediate formulas which can be useful
in reproducing our results and related calculations.

\subsection{One--Loop Integrals}

We define the relevant one--loop integrals as
\begin{equation}
(K,K_\rho,K_{\rho\sigma})=\int{d^D k\over(2\pi)^D}
{(1,k_\rho,k_\rho k_\sigma)\over
[(k-p_1)^2-M^2][(k+p_2)^2-\mu^2][k^2-\lambda^2]},
\end{equation}
where $D$ is the space--time dimension. Throughout the appendixes, it is
always assumed that $p_1^2=M^2\gg p_2^2=\mu^2 \gg \lambda^2$.
$M\,$($\mu$) plays the r\^{o}le of the b or c (u or d) quark
mass, whereas $\lambda$ is the gluon mass which along with $\mu$ is
used to regularize possible IR singularities of the integrals. The
renormalization scale appearing in regularized UV divergent integrals
is denoted by $\nu$.

The one--loop integrals can be parametrized as
\begin{eqnarray}
K&=&-{i\over(4\pi)^2}\,\widetilde{K}(M^2,q^2),\qquad q=p_1+p_2,\\
K_\rho&=&-{i\over(4\pi)^2}\left\{-2B(M^2,q^2) q_\rho-2\widetilde{B}
(M^2,q^2){p_2}_\rho\right\},\\
K_{\rho\sigma}&=&-{i\over(4\pi)^2}\left\{
A(M^2,q^2) q_\rho q_\sigma +B(M^2,q^2)[{p_2}_\rho
q_\sigma+{p_2}_\sigma q_\rho]
+\widetilde{B}(M^2,q^2){p_2}_\rho{p_2}_\sigma\right.\nonumber\\
&&{}+\left. C(M^2,q^2)
 g_{\rho\sigma}\right\}
\end{eqnarray}
with
\begin{eqnarray}
\widetilde{K}(M^2,q^2)&=&{1\over4(M^2-q^2)}\left\{
2\ln{\lambda^2\over M^2-q^2}\left( \ln{\mu^2\over M^2-q^2}+
\ln{M^2\over M^2-q^2}\right)\right.\nonumber\\
&&{}-\left.
\ln^2{\mu^2\over M^2-q^2}-\ln^2{M^2\over M^2-q^2}
-4\,{\rm L}_2\left(q^2\over q^2-M^2 \right)\right\},\\
A(M^2,q^2)&=&{1\over2q^2}\left\{1+\left(1-{M^2\over q^2}\right)
\ln{M^2\over M^2-q^2}\right\},\\
B(M^2,q^2)&=&-{1\over2q^2}\ln{M^2\over
M^2-q^2},\\
\widetilde{B}(M^2,q^2)&=&-{1\over2(M^2-q^2)}\left\{
\ln{\mu^2\over M^2-q^2}+\ln{M^2\over
M^2-q^2}\right\},\\
C(M^2,q^2)&=&-{\nu^{D-4} \over4}\left\{{2\over4-D}+\ln{4\pi}-
\gamma_E-\ln{M^2\over \nu^2}\right.\nonumber\\
&&{}+\left.
3+\left(1-{M^2\over q^2}\right)\ln{M^2\over
M^2-q^2}
\right\}.
\end{eqnarray}
Only $C(M^2,q^2)$ is UV divergent. Both $\widetilde{K}(M^2,q^2) $ and
$\widetilde{B}(M^2,q^2)$ are IR divergent, while $A(M^2,q^2)$ and
$B(M^2,q^2) $ are UV and IR finite.

For $q^2>M^2$ these functions become complex. It then proves
convenient to separate real and imaginary parts explicitly:
\begin{eqnarray}
\widetilde{K}(M^2,q^2>M^2)&=&{1\over4(M^2-q^2)}\left\{
2\ln{\lambda^2\over q^2-M^2}\left( \ln{\mu^2\over q^2-M^2}+
\ln{M^2\over q^2-M^2}\right)\right.\nonumber\\
&&{}-
\ln^2{\mu^2\over q^2-M^2}-\ln^2{M^2\over q^2-M^2}
+4\,{\rm L}_2\left(M^2\over M^2-q^2 \right)\nonumber\\
&&{}+\left. 4\ln{q^2\over q^2-M^2}\ln{M^2\over q^2-M^2}-{8\over3}\pi^2
\right\}\nonumber\\
&&{}+{i\pi\over M^2-q^2}\left(\ln{\lambda^2\over q^2-M^2}+
\ln{q^2\over q^2-M^2}\right),\\
A(M^2,q^2>M^2)&=&{1\over2q^2}\left\{1+\left(1-{M^2\over
q^2}\right)\left(\ln{M^2\over q^2-M^2}+i\pi\right)\right\},\\
B(M^2,q^2>M^2)&=&-{1\over2q^2}\left(\ln{M^2\over
q^2-M^2}+i\pi\right),\\
\widetilde{B}(M^2,q^2>M^2)&=&-{1\over2(M^2-q^2)}\left\{
\ln{\mu^2\over q^2-M^2}+\ln{M^2\over
q^2-M^2}+2i\pi\right\},\\
C(M^2,q^2>M^2)&=&-{\nu^{D-4} \over4}\left\{{2\over4-D}+\ln{4\pi}-
\gamma_E-\ln{M^2\over \nu^2}\right.\nonumber\\
&&{}+\left.
3+\left(1-{M^2\over q^2}\right)\left(\ln{M^2\over
q^2-M^2}+i\pi\right)
\right\}.
\end{eqnarray}

\subsection{Phase--Space Integrals}
Here we collect only the most involved three--particle
phase--space integrals. The simpler ones, as well as formulas for
$n$--particle phase--space integrals can be found in~\cite{Pi83}.

We define
\begin{equation}
 \int \mbox{\rm LIPS}(p_1,p_2,k) \equiv \int
{d^3 {p_1}\over2E_1}\,{d^3 p_2\over2E_2}\,{d^3 k\over2E_k}\,
\delta^4(P-p_1-p_2-k)
\end{equation}
 with $p_2^2=\mu^2$, $k^2=\lambda^2$ and keep only logarithms of
$\lambda^2$ and $\mu^2$.

\begin{eqnarray}
 \int {\mbox{\rm LIPS}(p_1,p_2,k)\over (\lambda^2 -2 P\cdot k)
(\lambda^2+2p_2\cdot k)} & = &
{}-{\pi^2\over4 P^2}\left\{ (P^2-p_1^2)\widetilde{K}(P^2,p_1^2)
+E(P^2,p_1^2)\right\},\label{ph-sp1}\\
 \int {\mbox{\rm LIPS}(p_1,p_2,k)\over \lambda^2 -2 P\cdot k}
&=&{}-{\pi^2\over4 P^2}\left\{
P^2-p_1^2-p_1^2\ln{P^2\over p_1^2}\right\},\\
 \int {\mbox{\rm LIPS}(p_1,p_2,k)
\,2 p_2\cdot k\over \lambda^2 -2 P\cdot k}&=&{}-{\pi^2\over16 P^2}
\left\{(P^2-p_1^2)(P^2+5p_1^2)
- 2p_1^2(p_1^2+2P^2)\ln{P^2\over p_1^2}\right\},\nonumber\\[-8pt]
& &\\[-20pt]
 \int {\mbox{\rm LIPS}(p_1,p_2,k)\over \lambda^2 +2 p_2\cdot k}
&=&-{\pi^2\over4 P^2}\left\{
 H(P^2,p_1^2)-2(P^2-p_1^2)^2 \widetilde{B}(P^2,p_1^2) \right\}
\nonumber\\
&=&{}-{\pi^2\over4 P^2}\left\{
 2(P^2-p_1^2)^2 {\rm Re}\,\widetilde{B}(p_1^2,P^2)-H(p_1^2,P^2)\right\},
\label{ph-sp2}\\
 \int {\mbox{\rm LIPS}(p_1,p_2,k)
\over \lambda^2 +2 p_1\cdot k}&=&{}-{\pi^2\over4 P^2}\left\{
P^2-p_1^2+P^2\ln{p_1^2\over P^2}\right\},\\
 \int {\mbox{\rm LIPS}(p_1,p_2,k)\over(\lambda^2+2p_1\cdot k)
(\lambda^2+2p_2\cdot k)}&=&{}-{\pi^2\over4 P^2}
\left\{(P^2-p_1^2){\rm Re}\,\widetilde{K}(p_1^2,P^2)+E'(p_1^2,P^2)
\right\},\label{ph-sp3}\\
\lefteqn{\kern-5.3cm\int \mbox{\rm LIPS}(p_1,p_2,k)\,
{2(p_1+p_2)\cdot k\over(\lambda^2+2p_1\cdot k)(
\lambda^2+2p_2\cdot k)}=}\nonumber\\
&=&{}-{\pi^2\over2 P^2}\left\{(P^2-p_1^2)^2{\rm Re}\,
\widetilde{B}(p_1^2,P^2)+P^2-p_1^2\right\}.
\label{ph-sp4}
\end{eqnarray}
In addition to the functions defined in the last
subsection, we also have used
\begin{eqnarray}
E(x,y)&=&\ln{y\over x-y}\ln{x\over x-y}-{\pi^2\over3}+
2\,{\rm L}_2\left({y\over y-x}\right)\nonumber\\
&=&\ln{y\over x}\ln{x\over x-y}-{\pi^2\over3}-
2\,{\rm L}_2\left({y\over x}\right),\\
E'(x,y)&=&E(y,x)+{\pi^2\over3},\\
H(x,y)&=&x-y+y\ln{x\over y}.
\end{eqnarray}
Note that (\ref{ph-sp1}), (\ref{ph-sp2}), (\ref{ph-sp3})
and~(\ref{ph-sp4}) are IR divergent.

\subsection{Integrals Related to Polylogarithms}
Our approach yields the relevant diagrams in terms of an integral over
the invariant mass squared
of two quarks, either both massless or one massive and
the other one massless. Using the formulas given in this appendix,
the remaining integral can also be done. There appear at
most polylogarithms, which are defined as
\begin{equation}
{\rm L}_n(y)=\int_0^y {dx\over x}\,{\rm L}_{n-1}(x),\qquad
{\rm L}_0\equiv -\ln(1-y),
\end{equation}
and have the following simple series expansion in the unit disc:
\begin{equation}
{\rm L}_n(y)=\sum_{k=1}^\infty {y^k\over k^n},\qquad |y|<1.
\end{equation}
We need the generic integrals
\begin{eqnarray}
I_n&=&\int_a^1{\rm d}x\, x^n \ln(1-x)\,\ln x,\\
J_n&=&\int_a^1{\rm d}x\, x^n {\rm L}_2(x),\\
K_n&=&\int_a^1{\rm d}x\, x^n \ln(x-a)\ln x,\\
L_n&=&\int_a^1{\rm d}x\, x^n {\rm L}_2\left({a\over x}\right)
\end{eqnarray}
for integer $n$. They can be computed with the help of the following
recursion formulas :
\begin{eqnarray}
I_n(n>0)&=&{n\over n+1}\, I_{n-1} +
   {a^n \ln a\over n+1}\left\{a+(1-a)\ln(1-a) \right\}\nonumber\\
   &&{}+{1\over n+1}\int_a^1{\rm d}x\left\{
         x^n + x^{n-1}(1-x)\ln(1-x)+n\, x^n \ln x
              \right\},\\
I_0&=&2-2a-{\pi^2\over6}-(1-a)(1-\ln a)\ln(1-a)+a\ln a
+{\rm L}_2(a),\\
I_{-1}&=&{\rm L}_2(a)\ln a +\zeta(3)-{\rm L}_3(a),\\
I_n(n<-1)&=&{1\over n+1}\left\{
-a^{n+1}\ln(1-a)\ln a -{\rm L}_2(1-a)-{1\over2}\ln^2 a
\right\}\nonumber\\
&&{}-{1\over n+1}\sum_{k=2}^{-n-1} {1\over(k-1)^2}
\left\{ 1-a^{1-k}
\left[ 1+(k-1)\ln a\right]
       \right\}\nonumber\\
& & {}-{1\over n+1}\int_a^1 {{\rm d}x\over x^{-n}}\ln(1-x),
\label{eq:Inneg}\\
J_n(n\neq -1)&=&{1\over n+1}\left\{
{\pi^2\over6}-a^{n+1}{\rm L}_2(a)+\int_a^1{\rm d}x\, x^n
\ln(1-x)\right\},\\
J_{-1}&=&\zeta(3)-{\rm L}_3(a),\\
K_n&=&a^{n+1}\int_a^1{{\rm d}x\over x^{n+2}}
\left\{\ln^2\left(x\over a\right)+\ln(1-x)\ln a \right\}-a^{n+1}
I_{-n-2}, \\
L_n&=&a^{n+1} J_{-n-2}.
\end{eqnarray}
Note that the sum in (\ref{eq:Inneg}) evaluates to zero for $n=-2$.

\section*{Appendix B}
\setcounter{equation}{0}
\addtocounter{section}{1}
\setcounter{subsection}{0}

In this appendix we give some details of the computation of the
imaginary parts of the diagrams II, V, VI, VIII, X, XI and XII. We have
used Cutkosky rules, hence, we think it may be useful
for the reader who wishes to reproduce our calculations to give the
contributions of the different cuts separately.

Colour--factors are omitted throughout the appendix. In order to get the
final answers for the diagrams, one has to multiply by the appropriate
colour--factors shown in Tab.~\ref{tab:1}.

We have used the NDR scheme with anticommuting $\gamma_5$; for a
discussion of this procedure cf.\ Sec.~2.3. We work in Feynman gauge
and use $\overline{\mbox{MS}}$ subtraction. In contrast to App.~A, the
renormalization scale is now denoted by $\mu$. The definition of the b
quark mass as pole mass is fixed by the treatment of the diagrams II and
XII. The c quark mass is defined by the treatment of the
graphs containing the c quark self--energy; we likewise choose the
pole mass definition.

Throughout this appendix, we use $a=(m_c/m_b)^2$. The phase--space
factor $f_1$ was defined in (\ref{eq:defPS}).

\subsection{Diagram II}

Diagram II is the simplest one. One only needs to multiply the lowest
order diagram I by the finite part of the b quark's on--shell
wave--function renormalization constant, $Z_{{\rm 2F}}$:
\begin{equation}
 Z_{\rm 2F} ={1\over 1-\Sigma'(m_b)}=1+C_F\,{g_s^2\over (4\pi)^2}
\left(\frac{2}{D-4}+ \gamma_E-\ln 4\pi +
\ln{m_b^2\over\mu^2}-2\ln{\lambda^2\over m_b^2}-4\right),
\end{equation}
where $\Sigma'(m_b)$ is the derivative of the b quark self--energy
with respect to $p\kern-0.2cm/$ at $p\kern-0.2cm/=m_b$.
The UV divergent piece of $Z_{{\rm 2F}}$ can be subtracted
from the imaginary part of diagram II before performing the
phase--space integrals. One finally gets (neglecting the
colour--factor $C_F$):
\begin{equation}
{\rm Im}\,[{\rm II+II}^\dagger]
={g_s^2 m_b^2\over 1536\pi^5}\left(\ln{m_b^2\over\mu^2}-
2\ln{\lambda^2\over m_b^2}-4\right) f_1(a).
\end{equation}
The IR divergence in the gluon mass, $\ln \lambda^2/m_b^2$, is an
artifact of the on--shell normalization and cancels against a
corresponding term in ${\rm Im}\,{\rm XII}$.

\subsection{Diagram V}

This diagram involves the d quark self--energy. It can be computed in
several ways of which we sketch only the simplest one. We first
observe that the diagrams IV and V are equal, since the u and the d
quark have equal masses, which immediately allows
us to write a relation analogous to (\ref{eq:anarel}) for the twice
Fierz--transformed diagram:
\begin{equation}\label{eq:abrakadabra}
{\rm Im}\,{\rm V} = \frac{1}{4\pi m_b^2}\,\int_{m_c^2}^{m_b^2}\!\!
ds\, (m_b^2-s)^2\left[ m_b^2\,\rho_1^{\rm SEL}(s) - 2 \rho_2^{\rm SEL}(s)
\right]
\end{equation}
where the $\rho_i^{\rm SEL}$ are the spectral densities of the light
quark's self--energy contribution to the vector correlation function
$\Pi_{\mu\nu}$, Eq.\ (\ref{eq:generalis}). They are given in Eqs.\
(\ref{eq:rhoSEL1}) and (\ref{eq:rhoSEL2}).
Performing the integral in (\ref{eq:abrakadabra}) then yields:
\begin{eqnarray}
{\rm Im}\,{\rm V} &=&  {g_s^2 m_b^6\over18432\pi^5}
\left\{
(1-a^2)(-43 +296 a-43 a^2)-48 a^2 \, \pi^2
\phantom{\ln{m_b^2\over\nu^2}} \right. \nonumber\\
 &&{}+  12 f_1(a) \left[2\ln(1-a)+\ln{m_b^2\over\nu^2}
\right]+288 a^2\, \ln a\;\ln(1-a)
\nonumber\\
 &&{}+ \left. 12 a^2(18-8a+a^2)\ln a +
 288 a^2\,{\rm L}_2(a)\right\}.
\end{eqnarray}

\subsection{Diagram VI}

Applying Cutkosky rules and using the formulas of App.~A,
both three--particle cuts yield a complex
result. In the sum, however, only the real parts contribute, for the
cuts are complex conjugate to each other. One finds:
\begin{eqnarray}
{\rm Im}\,{\rm VI}^{\mbox{\scriptsize (3--part.)}}&=&
-{g_s^2\over192\pi^5 m_b^2}\int_{m_c^2}^{m_b^2}{ds\over s^3}
(s-m_c^2)^2(m_b^2-s)^2\left\{
\left[2s^2+(m_c^2+m_b^2)s+2m_c^2 m_b^2 \right]\right. \nonumber\\
&&{}\times \left\{ (s-m_c^2)\left[2{\rm Re}\,\widetilde{B}(m_c^2,s)
-{\rm Re}\,\widetilde{K}(m_c^2,s)\right]+2{\rm Re}\,C(m_c^2,s)+1\right\}
\nonumber\\
&&{}+s\left[2s^2+(m_b^2+m_c^2)s-m_c^2 m_b^2\right]{\rm Re}\,A(m_c^2,s)
\nonumber\\
&&{}+ \left. s\left[2s^2+(4m_c^2+m_b^2)s+5m_c^2 m_b^2\right]{\rm Re}\,
B(m_c^2,s) \right\}.
\end{eqnarray}
The pole of the function $C(m_b^2,s)$ in $D-4$ cancels after addition
of the counterterm diagram. Hence, only the UV finite pieces must be
retained. Yet the subtraction of UV divergences has to be done with
some care: counterterm diagrams must be added {\em before}
performing the phase--space integration, which
we explicitly carry out in {\em four} dimensions. This procedure has the
advantage that one can ignore the evanescent operators.
Note however that the three--particle cuts are IR divergent since they
contain the
functions $\widetilde{K}(m_b^2,s)$ and $\widetilde{B}(m_b^2,s)$.

With the help of App.~A one can also check that the sum of the two
four--particle cuts is
\begin{eqnarray}
{\rm Im}\,{\rm VI}^{\mbox{\scriptsize (4--part.)}}&=&
-{g_s^2\over192\pi^5 m_b^2} \int_{m_c^2}^{m_b^2}{ds\over s^3}
(m_b-s)^2\left\{
\left[2s^2+(m_c^2+m_b^2)s+2m_c^2 m_b^2 \right]\phantom{\ln{s\over m_c^2}}
\kern-2.0cm \right. \nonumber\\
&&{}\times(s-m_c^2)^2\left\{
(s-m_c^2)\left[
{\rm Re}\,\widetilde{K}(m_c^2,s)-2{\rm Re}\,\widetilde{B}
(m_c^2,s)\right]+E'(m_c^2,s)
\right\}\nonumber\\
&&{}-m_c^2\left[s^2+(2m_c^2-m_b^2)s+4m_c^2m_b^2\right] H(m_c^2,s)
\nonumber\\
&&{}-{sm_c^2\over2}
(s+2m_c^2)(s+2m_b^2)\ln{s\over m_c^2}-{s-m_c^2\over12}\left[46s^3\right.
\nonumber\\
&&\left.\left.{}-(11m_c^2-20m_b^2)s^2+
m_c^2(19m_c^2-22m_b^2)s+38m_c^4m_b^2
\right]
\phantom{\ln{s\over m_c^2}}
\kern-1.2cm \right\}.
\end{eqnarray}
Obviously, the four--particle cuts are UV finite, because they do not
involve any loop integration. Note that the IR divergences
cancel in the sum of the three-- and four--particle cuts.
 The result can be integrated with the help of App.~A to yield:
\begin{eqnarray}
{\rm Im}\,{\rm VI}&=&{g_s^2 {m_b}^6 \over 9216 \pi^5} \left\{
(1- a) (61 - 165 a - 390 a^2 + 52 a^3)
+ 4 a^2 (30 + 8 a - a^2)\pi^2 \kern-2.0cm
\phantom{\ln\left({{m_b}^2\over \mu^2}\right)}  \right.\nonumber \\
& &{}- 2 (1 - a^2) \left[25 - 272 a + 25 a^2
-12(1 - 8 a + a^2)\ln a\right] \ln(1 - a) \nonumber\\
 & &{}+2 a (12 + 18 a + 212 a^2 - 19 a^3) \ln a
- 12 a^2 (18 + 8 a - a^2) \ln^2 a
- 12 f_1(a)
\ln{{m_b}^2\over \mu^2}\makebox[4mm]{ } \nonumber\\
& &\left.{}+ 24 (2 - 16 a - 30 a^2 + 8 a^3 - a^4 + 12 a^2 \ln a)
 {\rm L}_2(a) +
 864a^2\! \left[\zeta(3)-{\rm L}_3(a) \right]\right\}\!.
\label{ApBVI}
\end{eqnarray}

\subsection{Diagram VIII}

Proceeding with diagram~VIII along the same lines, one gets for the
sum of the two three--particle cuts:
\begin{eqnarray}
{\rm Im}\,{\rm VIII}^{\mbox{\scriptsize (3--part.)}}&=&
{g_s^2\over32\pi^5 m_b^2}
\int_{m_c^2}^{m_b^2}{ds\over s}(m_b^2-s)^2(s-m_c^2)^2
\left\{ 2s\left[{\rm Re}\,A(m_c^2,s)+{\rm Re}\,B(m_c^2,s)\right]
\phantom{{1\over2}}\kern-0.7cm \right.\nonumber\\
&&\left.{}+ (s-m_c^2)\left[2\,{\rm Re}\,\widetilde{B}(m_c^2,s)-
{\rm Re}\,\widetilde{K}(m_c^2,s)\right]+ 8\,{\rm Re}\,C(m_c^2,s)
+{1\over2}\right\}\!,\makebox[1.2cm]{ }
\end{eqnarray}
while the sum of the two four--particle cuts is
\begin{eqnarray}
{\rm Im}\,{\rm VIII}^{\mbox{\scriptsize (4--part.)}}
&=&
-{g_s^2\over32\pi^5 m_b^2}
\int_{m_c^2}^{m_b^2}{ds\over s}(m_b^2-s)^2
\left\{(s-m_c^2)^3 \left[2{\rm Re}\,\widetilde{B}(m_c^2,s)-
{\rm Re}\,\widetilde{K}(m_c^2,s)\right] \phantom{{5s-m_c^2\over2}}
\kern-2.0cm\right.\nonumber\\
&&{}- \left. (s-m_c^2)^2 E'(m_c^2,s)
+(s-m_c^2){5s-m_c^2\over2}-2m_c^2s\ln{s\over m^2_c}
\right\}.
\end{eqnarray}
Again, UV divergences are cancelled by the counterterm diagrams, whereas
the IR divergences add to zero in the final sum. The latter one can
be integrated to give:
\begin{eqnarray}
{\rm Im}\,{\rm VIII}&=&{g_s^2 m_b^6 \over 9216 \pi^5} \left\{
(1 - a) (-220 + 1374 a + 1863 a^2 - 139 a^3) -
4 a^2 (66 + 32 a - a^2)\pi^2
\vphantom{\ln\left({{m_b}^2 \over \mu^2}\right)}\right.\nonumber \\
& &{} + 2 (1 - a) \left[37 - 379 a - 811 a^2 - 11 a^3
-12(1 + a) (1 - 8 a + a^2)\ln a\right] \ln(1 - a)  \nonumber \\
& &{} - 2 a (12 - 918 a + 596 a^2 + 35 a^3) \ln a
+12 a^2 (18 + 32 a - a^2) \ln^2 a
 \nonumber \\
& & {}-24(2 - 16 a - 66 a^2 - 16 a^3 - a^4 + 12 a^2 \ln a )\,
{\rm L}_2(a)\nonumber\\
& & \left.{}+48 f_1(a) \ln{{m_b}^2 \over \mu^2} +
864 a^2 \left[{\rm L}_3(a) -
\zeta(3)\right]\vphantom{\frac{g_s^2 m_b^6}{9216 \pi^5}}\right\}.
\label{ApBVIII}
\end{eqnarray}

\subsection{Diagram X}

With the same notations as above, we get for diagram~X the following
results:
\begin{eqnarray}
{\rm Im}\,{\rm X}^{\mbox{\scriptsize (3--part.)}}&=&
{g_s^2\over64\pi^5 m_b^2}
\int_{m_c^2}^{m_b^2}{ds\over s}\,(m_b^2-s)^2(s-m_c^2)^2\left\{
2s[A(m_b^2,s)+B(m_b^2,s)]\phantom{{1\over2}}\right.\nonumber\\
&&{}+\left. 8 C(m_b^2,s)+{1\over2}+(m_b^2-s)[\widetilde{K}(m_b^2,s)\!
-\!2\widetilde{B}(m_b^2,s)]\right\}.
\end{eqnarray}
Note that $A(m_b^2,s)$, $B(m_b^2,s)$, $C(m_b^2,s)$,
$\widetilde{K}(m_b^2,s)$ and $\widetilde{B}(m_b^2,s)$ are now real
functions because $s<m_b^2$. By adding the counterterm diagram one
removes the UV divergent piece in~$C(m_b^2,s)$, as it should be.

The four--particle cut yields
\begin{eqnarray}
{\rm Im}\,{\rm X}^{\mbox{\scriptsize (4--part.)}}
&=&
-{g_s^2\over64\pi^5 m_b^2}
\int_{m_c^2}^{m_b^2}{ds\over s}(s-m_c^2)^2
\left\{(m_b^2-s)^3 \left[\widetilde{K}(m_b^2,s)-
2\widetilde{B}(m_b^2,s)
\right]\phantom{\ln{m_b^2\over s}} \right.\nonumber\\
&&{}+ (m_b^2-s)^2 E(m_b^2,s)+(m_b^2-s) H(m_b^2,s)
-(m_b^2-s){3s+m_b^2\over2}\nonumber\\
&&{}+\left.s(s+m_b^2)\ln{m_b^2\over s}
\right\},
\end{eqnarray}
so that
\begin{eqnarray}
{\rm Im}\,[{\rm X}+{\rm X}^\dagger]&=&{g_s^2 {m_b}^6 \over 9216 \pi^5}
\left\{ (1 - a) (-139 + 1863 a + 1374 a^2 - 220 a^3)\right.\nonumber \\
& &{}+4 (3 - 48 a - 66 a^2 + 16 a^3 - 2 a^4){\pi ^2}
- 864 a^2 \left[{\rm L}_3(a) - \zeta(3)\right] -2 (1 - a) \times
\nonumber \\
& &{}\times\!\left[11 + 811 a + 379 a^2 - 37 a^3
-12(1 + a) (1 - 8 a + a^2)  \ln a\right] \ln(1 - a)  \nonumber \\
& &{}+2 a (12 + 486 a - 236 a^2 + 13 a^3 - 48 a \pi^2) \ln a
+48 f_1(a) \ln{{m_b}^2\over \mu^2} \nonumber \\
& &\left.{}+
24 (1 + 16 a + 66 a^2 + 16 a^3 - 2 a^4 + 12 a^2 \ln a )\, {\rm L}_2(a)
\right\}\!.
\label{ApBX}
\end{eqnarray}
Again the result is both UV and IR finite as it should be.

\subsection{Diagram XI}

\begin{eqnarray}
{\rm Im}\,{\rm XI}^{\mbox{\scriptsize (3--part.)}}
&=&{g_s^2\over384\pi^5 m_b^2} \int_{m_c^2}^{m_b^2}{ds\over s^3}
(m_b^2-s)^2(s-m_c^2)^2\left\{
\left[2s^2+(m_c^2+m_b^2)s+2m_c^2 m_b^2 \right]\right. \nonumber\\
&&{}\times \left\{ (m_b^2-s)\left[2\widetilde{B}(m_b^2,s)
-\widetilde{K}(m_b^2,s)\right]-2C(m_b^2,s)-1\right\} \nonumber\\
&&{}-s\left[2s^2+(m_b^2+m_c^2)s-m_c^2 m_b^2\right]A(m_b^2,s)
\nonumber\\
&&{}- \left. s\left[2s^2+(m_c^2+4m_b^2)s+5m_c^2 m_b^2\right]B(m_b^2,s)
 \right\}\!.
\end{eqnarray}
$A(m_b^2,s)$, $B(m_b^2,s)$, $C(m_b^2,s)$,
$\widetilde{K}(m_b^2,s)$ and $\widetilde{B}(m_b^2,s)$ are real functions
and the counterterm diagram cancels the pole of $C(m_b^2,s)$ in $D-4$.

The four--particle cut is:
\begin{eqnarray}
{\rm Im}\,{\rm XI}^{\mbox{\scriptsize (4--part.)}}&=&
{g_s^2\over384\pi^5 m_b^2}
\int_{m_c^2}^{m_b^2}{ds\over s^3}
(s-m_c^2)^2\left\{
\left[2s^2+(m_c^2+m_b^2)s+2m_c^2 m_b^2 \right](m_b^2-s)
\phantom{\ln{m_b^2\over s}}\right. \nonumber\\
&&{}\times\!\left\{
(m_b^2-s)^2\left[
\widetilde{K}(m_b^2,s)-2\widetilde{B}(m_b^2,s)\right]+H(m_b^2,s)
+(m_b^2-s)E(m_b^2,s)
\right\}\nonumber\\
&&{}+{s^2\over2}
\left[4s^2+(2m_c^2-m_b^2)s-2m_c^2m_b^2\right]\ln{m_b^2\over s}
-{m_b^2-s\over12}\left[22s^3\right.\nonumber\\
&&{}\left.\left.{}+(8m_c^2-11m_b^2)s^2-m_b^2(22m_c^2-7m_b^2)s+
14m_c^2m_b^4\right]\right\}\!.
\end{eqnarray}
After integrating over $s$, the final sum can be written as:
\begin{eqnarray}
\lefteqn{
{\rm Im}\,[{\rm XI}+{\rm XI}^\dagger]= {g_s^2 {m_b}^6 \over 9216 \pi^5}
\left\{  (1 - a) (52 - 390 a - 165 a^2 + 61 a^3)\right.}
\nonumber\\&&{} -
 4 (3 - 24 a - 30 a^2 + 16 a^3 - 2 a^4)\pi^2
\phantom{\ln\left({{m_b}^2\over\mu^2}\right)}\kern-2.0cm
 \nonumber \\
&&{}- 2 (1 - a^2) \left[25 - 272 a + 25 a^2
+12(1 - 8 a + a^2)  \ln a\right] \ln(1 - a)  \nonumber \\
&&{}- 2 a (12 - 18 a - 236 a^2 + 19 a^3 - 48 a \pi^2) \ln a
- 12 f_1(a) \ln{{m_b}^2\over\mu^2}\nonumber\\
&&{}+ \left.
24 (-1 + 8 a - 30 a^2 - 16 a^3 + 2 a^4 - 12 a^2 \ln a )
\,{\rm L}_2(a) +  864 a^2 \left[
{ \rm L}_3(a) - \zeta(3)\right] \right\},\makebox[1cm]{ }
\label{ApBXI}
\end{eqnarray}
which again is explicitly UV and IR finite.

\subsection{Diagram XII}

For diagram~XII only a four--particle cut is possible. Using the same
techniques as before one finds:
\begin{eqnarray}
{\rm Im}\,{\rm XII} &=& -{g_s^2 \over 64 \pi^3 m_b^2}\int_{m_c^2}^{m_b^2}
ds\, \left\{
\left[m_b^2\rho^{(1)}_1(s)-2\rho^{(1)}_2(s)\right] \left[4 (m_b^2-s)^2
\vphantom{{m_b^2-s \over m_b^2}}\right. \right.\nonumber \\
&&\times\left.\left\{\ln  {m_b^2-s \over m_b^2}
-\frac{1}{2}\,\ln\, \frac{\lambda^2}{m_b^2}\right\} +
s(2 m_b^2 + 5s)
\ln {m_b^2 \over s} -{1 \over 2}(m_b^2-s)(17 m_b^2-3 s)\right]
\nonumber\\
& &\left.{}+ \rho^{(1)}_1(s)\left[-2s m_b ^2 (m_b ^2 + 3s)
\ln {m_b ^2 \over s} +{m_b ^2 -s \over 3} (s^2+28s m_b ^2 -5 m_b^4)
\right]
\right\}\nonumber\\
&=& {g_s^2 m_b^6\over18432\pi^5}\Big\{
137-1024 a-324 a^2+1216 a^3-5a^4+96 a^2\;\pi^2\nonumber\\
& &{}
-24 f_1(a)\left[2\ln(1-a)-\ln{\lambda^2\over m_b^2}\right]
\nonumber\\
&&{}- 12 a^2(102+8 a+5 a^2)\ln a-576 a^2
\left[\ln a\;\ln(1-a)+{\rm L}_2(a)\right]\Big\}.
\end{eqnarray}
The IR divergent logarithm of $\lambda$ cancels the IR divergence
in the sum of the diagrams~II and~II${}^\dagger$, thus rendering an IR
finite result.
Similarly, the dependence on the renormalization scale $\mu$ cancels
in the sum ${\rm II}+{\rm II^\dagger}+{\rm V}+{\rm XI}+{\rm
XI}^\dagger+{\rm XII}$, as it should be for the semileptonic decay rate
$\Gamma(b\to u\tau\bar\nu)$.

It is important to emphasize that these results do not change under
Fierz--trans\-for\-ma\-tions. This has been checked explicitly for the
seven diagrams discussed in this appendix.

\subsection{The Limit $m_c\to0$}\label{ss:mc}

Here, we show that our results are well--defined in the limit
$m_c\to0$ and compare them with the corresponding quantities obtained
in \cite{GB93}. From~(\ref{ApBVI}), (\ref{ApBVIII}),
(\ref{ApBX}) and~(\ref{ApBXI}) we get:
\begin{eqnarray}
\left.{\rm Im}\,{\rm VI}\,\right|_{m_c=0}&=&{g_s^2 m_b^6\over768}
\left({61\over12}
                       -\ln{m_b^2\over\mu^2}\right), \\
\left.{\rm Im}\,{\rm VIII}\,\right|_{m_c=0}&=&
{g_s^2 m_b^6\over768}\left(-{55\over3}
                       +4\ln{m_b^2\over\mu^2}\right), \\
\left.{\rm Im}[{\rm X}+{\rm X}^\dagger]\,\right|_{m_c=0}&=&
{g_s^2 m_b^6\over768}\left(-{139\over12}+\pi^2
                       +4\ln{m_b^2\over\mu^2}\right), \\
\left.{\rm Im}[{\rm XI}+{\rm XI}^\dagger]\,\right|_{m_c=0}&=&
{g_s^2 m_b^6\over768}\left({13\over3}-\pi^2
                       -\ln{m_b^2\over\mu^2}\right).
\end{eqnarray}
In the notation of~\cite{GB93} this reads:
\begin{eqnarray}
G_g&=&{61\over12}
                       -\ln{m_b^2\over\mu^2}, \\
G_e&=&-{55\over3}
                       +4\ln{m_b^2\over\mu^2}, \\
G_d&=&-{139\over12}+\pi^2
                       +4\ln{m_b^2\over\mu^2}, \\
G_f&=&{13\over3}-\pi^2
                       -\ln{m_b^2\over\mu^2}.
\end{eqnarray}
These results are scheme--dependent. In order to compare with~\cite{GB93},
where the 't~Hooft--Veltman scheme is used,
we need to add the corresponding matching coefficients,
$B_d$, $B_e$, $B_f$, $B_g$
evaluated using NDR. Again in the notation of~\cite{GB93} one has:
\begin{equation}
B_d=B_e=5,\qquad B_f=B_g={1\over2}
\end{equation}
with $B = B_d+B_e+B_f+B_g = 11$, cf.\ (\ref{eq:B}).
The agreement with~\cite{GB93} for the scheme--independent
quantities $G_i+B_i$ is evident,
which provides a non--trivial check of
 our calculation.

\section*{Appendix C}
\setcounter{equation}{0}
\addtocounter{section}{1}

As mentioned in the text, the calculation
of  $G_b$ and $G_d$ is much simplified by using the results for the
spectral densities of the vector correlation function, $1/\pi\,
{\rm Im}\,\Pi_{\mu\nu} = q_\mu q_\nu\, \rho^V_1 + g_{\mu\nu}\, \rho^V_2$,
Eq.\ (\ref{eq:generalis}), which can be obtained from \cite{Gen90}
when the running mass is replaced by the pole mass; we have
also calculated the spectral densities directly by using Cutkosky rules.
Defining
\begin{equation}
\rho^V_i = 3\, (\rho_i^{(1)} + C_F\,\rho_i^{(2)}+\dots),
\end{equation}
where $\rho_i^{(1)}$ is the tree--level and $\rho_i^{(2)}$ the
next--to--leading order contribution, we find:
\begin{eqnarray}
\rho^{(1)}_1(s) & = & \frac{1}{12\pi^2}
\left(1-\frac{m_c^2}{s}\right)^2 \left(1+\frac{2m_c^2}{s}\right),\\
\rho^{(1)}_2(s) & = & {}-\frac{1}{12\pi^2}\,s
\left(1-\frac{m_c^2}{s}\right)^2 \left(1+\frac{m_c^2}{2s}\right),\\
\rho^{(2)}_1(s)&=&{g_s^2\over192\pi^4}\left\{
4 \left(1-{m_c^2\over s}\right)^2 \left(1+{2 m_c^2\over s}\right)
\left[
\ln{m_c^2\over s}\ln\left(1-{m_c^2\over s}\right)+2
{\rm L}_2\left({m_c^2\over s}\right)
\right]\right.\nonumber\\
& &{}-4\left(1+2{m_c^2\over s}-2 {m_c^4\over s^2}\right)\ln{m_c^2\over s}
-4\left(1-{m_c^2\over s}\right)^2\left(1+5 {m_c^2\over s}-{
3m_c^4\over2 s^2}\right)\ln{s-m_c^2\over m_c^2}\nonumber\\
&&\left.{}+3\left(1-{m_c^2\over s}\right)
\left(1+3{m_c^2\over s}-{16m_c^4\over 3s^2}\right)
\right\},\\
\rho^{(2)}_2(s)&=&-{g_s^2\over192\pi^4}s\left\{
4 \left(1-{m_c^2\over s}\right)^2 \left(1+ {m_c^2\over 2 s}\right)
\left[
\ln{m_c^2\over s}\ln\left(1-{m_c^2\over s}\right)+2
{\rm L}_2\left({m_c^2\over s}\right)
\right]\right.\nonumber\\
&-&4\left(1+{m_c^2\over s}\right)\left(1-{m_c^2\over 2s}\right)
\ln{m_c^2\over s}
-4\left(1-{m_c^2\over s}\right)^2\left(1+{5m_c^2\over 4s}\right)
\ln{s-m_c^2\over m_c^2}\nonumber\\
&+&\left.3\left(1-{m_c^2\over s}\right)
\left(1-{3m_c^2\over2 s}-{5m_c^4\over6 s^2}\right)
\right\}.
\end{eqnarray}
We repeat that the above spectral densities are expressed in terms of
on--shell quark masses.
In the limit $m_c\to0$ one obtains
\begin{equation}
\rho^{(2)}_2(s)\Big|_{\rm massless}=
- s\,\rho^{(2)}_1(s)\Big|_{\rm massless}=-{g_s^2\over 64\pi^4}\,s
\end{equation}
and $\Pi_{\mu\nu}$ becomes transverse, as it should.

Since for the calculation of diagram V, App.~B.2., the part of
$\rho_i^{(2)}$ corresponding to the light quark self--energy is needed
separately, we give in addition to the sum, $\rho_i^{(2)}$, also
the contributions of
the single diagrams. All the three graphs are calculated in Feynman
gauge and expressed in terms of the
{\em running $\overline{\mbox{\rm MS}}$ mass}
whose relation to the pole mass is given by
\begin{equation}
m^{{\rm pole}} = m_{\overline{\mbox{\scriptsize MS}}}(\mu)\left[ 1 +
C_F\,\frac{g_s^2}{4\pi^2}\,\left(1 - \frac{3}{4}\,\ln\,\frac{m^2}{\mu^2}
\right)\! \right].
\end{equation}
We denote the spectral density of the light quark's self--energy
diagram by $\rho_i^{\rm SEL}$, that of the heavy quark by $\rho_i^{\rm
SEH}$ and the gluon--exchange diagram by $\rho_i^{\rm EX}$.
\begin{eqnarray}
\rho_1^{\rm SEL}(s) & = & \frac{g_s^2}{1152\pi^4} \left[
-\frac{s-m_c^2}{s}\, \left\{ 11 + 17\,\frac{m_c^2}{s} - 34
\left(\frac{m_c^2}{s}\right)^2\right\}- 6 \ln\,\frac{s}{m_c^2}
\right.\nonumber\\
& & \left.{} + 6\left( \frac{s-m_c^2}{s}
\right)^2 \frac{2m_c^2+s}{s} \left\{ \ln\, \frac{m_c^2}{\mu^2} + 2
\ln\, \frac{s-m_c^2}{m_c^2}\right\}\!\right],\label{eq:rhoSEL1}\\
\rho_2^{\rm SEL}(s) & = & \frac{g_s^2}{2304\pi^4} \left[
(s-m_c^2) \left\{ 28 -5 \,\frac{m_c^2}{s} - 17
\left(\frac{m_c^2}{s}\right)^2\right\}+6(2s-3m_c^2)\,\ln
\frac{s}{m_c^2}\right.\nonumber\\
& & \left.{} - 6\left( \frac{s-m_c^2}{s}
\right)^2 (2m_c^2+s) \left\{ \ln\, \frac{m_c^2}{\mu^2} + 2
\ln\, \frac{s-m_c^2}{m_c^2}\right\}\!\right],
\label{eq:rhoSEL2}\\
\rho_1^{\rm SEH}(s) & = &\frac{g_s^2}{1152\pi^4} \left[
-\frac{s-m_c^2}{s}\, \left\{ 11 - 55\,\frac{m_c^2}{s} +362
\left(\frac{m_c^2}{s}\right)^2\right\}+30\ln\frac{s}{m_c^2}
\right.\nonumber\\
& & {}
+ 6\frac{s-m_c^2}{s}\,
\left\{ 1 +\,\frac{m_c^2}{s} +34
\left(\frac{m_c^2}{s}\right)^2\right\}
 \ln\,\frac{m_c^2}{\mu^2}
\nonumber\\
& & \left.{}  -24\left( \frac{s-m_c^2}{s}
\right)^2 \left(1+2\,\frac{m_c^2}{s}\right)
\ln\, \frac{s-m_c^2}{m_c^2} \right],\\
\rho_2^{\rm SEH}(s) & = & \frac{g_s^2}{2304\pi^4} \left[
(s-m_c^2) \left\{ 28 +157\,\frac{m_c^2}{s} +181
\left(\frac{m_c^2}{s}\right)^2\right\}-6(10s+3m_c^2)\,\ln
\frac{s}{m_c^2} \right.\nonumber\\
& & {} -6(s-m_c^2) \left\{ 2 +17\,\frac{m_c^2}{s} +17
\left(\frac{m_c^2}{s}\right)^2\right\}
\ln\, \frac{m_c^2}{\mu^2}\nonumber\\
& & \left.{} +24\left( \frac{s-m_c^2}{s}
\right)^2 (2s+m_c^2)
\ln\, \frac{s-m_c^2}{m_c^2} \right],\\
\rho_1^{\rm EX}(s) & = & \frac{g_s^2}{288\pi^4}\,\frac{s-m_c^2}{s}
\left[\frac{s-m_c^2}{s} \left\{ 2\left(5+7\frac{m_c^2}{s}\right)
-3\left[1+8\,\frac{m_c^2}{s}-3\left(\frac{m_c^2}{s}\right)^2\right]
\ln\,\frac{s-m_c^2}{m_c^2}\right.\right.\nonumber\\
& &\left. {}-3\left(1+2\,\frac{m_c^2}{s}\right)\!\left[\ln\,
\frac{m_c^2}{\mu^2}+2\ln\,\frac{s}{m_c^2}\,\ln\,\frac{s-m_c^2}{s}-
4{\rm L}_2\left(\frac{m_c^2}{s}\right)\right]\right\}
\nonumber\\
& &\left. {}+ 12\,\frac{m_c^2}{s}\,\ln\,\frac{s}{m_c^2}\right],\\
\rho_2^{\rm EX}(s) & = & \frac{g_s^2}{1152\pi^4} \left[
-(s-m_c^2)\left\{46-23\,\frac{m_c^2}{s}-5\left(\frac{m_c^2}{s}\right)^2
\right\}+6m_c^2\left(1+2\,\frac{m_c^2}{s}\right)\ln\,
\frac{s}{m_c^2} \right.\nonumber\\
& & {} +6\left(\frac{s-m_c^2}{s}\right)^2
\left\{2(s+2m_c^2)\,\ln\, \frac{s-m_c^2}{m_c^2}+
(2s+m_c^2)\left[\ln\,\frac{m_c^2}{\mu^2}\right.\right.\nonumber\\
& &\left.\left.\left. {} +2\ln\,\frac{s}{m_c^2}\,\ln\,\frac{s-m_c^2}{s}
-4{\rm L}_2\left(\frac{m_c^2}{s}\right)\right]\right\}\right].
\end{eqnarray}

\renewcommand{\textfraction}{0}
\clearpage

\section*{Figures}

\clearpage
\clearpage

\clearpage

\section*{Tables}
\begin{table}[h]
\addtolength{\arraycolsep}{0.2cm}
\renewcommand{\arraystretch}{1.3}
$$
\begin{array}{l|ccc}
\mbox{Diagram No.} & {\cal O}_1\otimes {\cal O}_1 & {\cal O}_1\otimes
{\cal O}_2 & {\cal O}_2\otimes {\cal O}_2\\ \hline
\mbox{I}   &  3       & 1   & 3      \\
\mbox{II}  &  3 \,C_F & C_F & 3 \,C_F\\
\mbox{III} &  3 \,C_F & C_F & 3 \,C_F\\
\mbox{IV}  &  3 \,C_F & C_F & 3 \,C_F\\
\mbox{V}   &  3 \,C_F & C_F & 3 \,C_F\\
\mbox{VI}  &  3 \,C_F & C_F & 0      \\
\mbox{VII} &  0       & C_F & 3 \,C_F\\
\mbox{VIII}&  0       & C_F & 0      \\
\mbox{IX}  &  0       & C_F & 3 \,C_F\\
\mbox{X}   &  0       & C_F & 0      \\
\mbox{XI}  &  3 \,C_F & C_F & 0      \\
\mbox{XII} &  3 \,C_F & C_F & 3 \,C_F
\end{array}
$$
\caption[]{The colour--factors multiplying the diagrams shown in
Fig.~\ref{fig:1} for all possible combinations of operator
insertions. For an arbitrary number of colours $N_c$, $C_F$ equals
$(N_c^2-1)/(2N_c)$.}\label{tab:1}
$$
\addtolength{\arraycolsep}{-0.1cm}
\begin{array}{c|llllll}
\sqrt{a} = m_c/m_b & \phantom{-}0 & \phantom{-}0.1 &
\phantom{-}0.2 & \phantom{-}0.3 & \phantom{-}0.4 & \approx 0.5\\ \hline
H_a(a) & -3.62 & -3.22 & -2.71 & -2.21 & -1.75\\
H_b(a) & \phantom{-}1.5 & \phantom{-}2.43 & \phantom{-}4.20 &
\phantom{-}6.70 & \phantom{-}10.5 & \frac{739}{70} +
\frac{2}{3}\,\pi^2-15\,\ln 2 - 3\,\ln \left(\frac{1}{4}-a\right)\\
H_c(a) & -3.62 & -3.58 & -3.44 & -3.11 & -2.16\\
H_d(a) & \phantom{-}1.5 & \phantom{-}3.37 & \phantom{-}6.46
& \phantom{-}12.8 & \phantom{-}24.3
\end{array}
\addtolength{\arraycolsep}{0.1cm}
$$
\caption[]{Numerical values of the functions $H_x$ defined in
Sec.~4.}\label{tab:3}
\end{table}
\clearpage
\begin{table}[h]
\renewcommand{\arraystretch}{1.3}
$$
\begin{array}{l|lllll}
\sqrt{a} = m_c/m_b & \phantom{-}0 & \phantom{-}0.1 & \phantom{-}0.2 &
\phantom{-}0.3 & \phantom{-}0.4 \\ \hline
\Gamma(b\to c\bar c s) & \phantom{-}1.112 & \phantom{-}1.005
& \phantom{-}0.671 & \phantom{-}0.283 & \phantom{-}0.043\\
\times 64\pi^3/(G_F^2m_b^5) & \pm 0 & \pm 0.006 & \pm 0.008 & \pm 0.005
& \pm 0.001
\end{array}
$$
\caption[]{The decay rate $\Gamma(b\to c\bar c s)$ in units of
$G_F^2m_b^5/(64\pi^3)$ as function of $m_c/m_b$ for $\mu = m_b$ and
$\alpha_s(m_Z) = 0.117$. The error denotes the uncertainty in the
radiative corrections to the free quark decay.}\label{tab:x}
$$
\begin{array}{c|lll}
 \sqrt{a} = m_c/m_b & \phantom{-}c_{11} & c_{12}(\mu =m_b) &
\phantom{-}c_{22}\\ \hline
0 & \phantom{-}{{31}\over 4} - \pi^2 & -{7\over 4} - \pi^2 &
\phantom{-}{{31}\over 4} - \pi^2 \\
 0.1 & -1.2 & -11. & -1.8 \\
 0.2 & \phantom{-}0.32 & -11. & -1.3 \\
 0.3 & \phantom{-}2.0 & -10. & -1.0 \\
 0.4 & \phantom{-}3.8 & -9.5 & -0.73 \\
 0.5 & \phantom{-}5.6 & -8.9 & -0.51 \\
 0.6 & \phantom{-}7.6 & -8.4 & -0.33 \\
 0.7 & \phantom{-}9.9 & -7.8 & -0.20 \\
 0.8 & \phantom{-}13. & -7.2 & -0.094 \\
 0.9 & \phantom{-}17. & -6.6 & -0.027 \\
 1 & \phantom{-}{{147}\over {10}} - {2\over 3}\,\pi^2 - 6\,\ln (1 - a)
& -6 & \phantom{-}0
\end{array}
$$
\caption[]{Numerical values of the coeffiencts $c_{ij}$ defined in
(\protect{\ref{eq:cij}}).}\label{tab:2}
$$
\begin{array}{c|ccc}
\alpha_s(m_Z) & \mbox{Parton Model \protect{\cite{AP91}}} & \mbox{HQE
\protect{\cite{BBSV941}}} & \mbox{HQE [this work]} \\
\hline
0.110 & 0.132 & 0.130 & 0.121\\
0.117 & 0.128 & 0.126 & 0.116\\
0.124 & 0.124 & 0.121 & 0.111
\end{array}
$$
\caption[]{$B(B\to Xe\nu)$ in different models depending on
$\alpha_s(m_Z)$. Input parameters: $m_b=4.8\,$GeV, $m_c = 1.3\,$GeV,
which corresponds to $\lambda_1 = -0.6\,$GeV$^2$. In the phase--space
factor of $\Gamma(b\to c\bar c s)$ we use $m_s=0.2\,$GeV. HQE is
short--hand for heavy quark expansion.}\label{tab:extra}
\end{table}

\end{document}